\documentclass[journal,10pt,twocolumn]{IEEEtran}

\usepackage{cite}
\usepackage{amssymb}
\usepackage{amsmath}
\usepackage{graphicx}
\usepackage{enumitem}
\usepackage{bm}
\usepackage{mathtools}
\usepackage{amsfonts}
\usepackage{stackengine}
\usepackage{xcolor}
\usepackage{stfloats}
\usepackage{array}
\newcolumntype{M}[1]{>{\centering\arraybackslash}m{#1}}
\newcolumntype{P}[1]{>{\centering\arraybackslash}p{#1}}
\stackMath
\usepackage{multirow} 
\usepackage{framed}
\newcommand{\kernel}{{\ooalign{$k$\cr\raisebox{0.2em}{\kern0.08em--}\cr}
}}
\DeclareRobustCommand{\mhl}[1]{%
	\ifmmode\text{\color{black}{$#1$}}\else{\color{black}{#1}}\fi
}
\makeatletter
\newcommand{\vast}{\bBigg@{3}}
\makeatother

\makeatletter
\@ifdefinable\@latex@chi{\let\@latex@chi\chi}
\renewcommand*\chi{{\@latex@chi\smash[t]{\mathstrut}}} 
\makeatletter
\def\mathclap#1{\text{\hbox to 0pt{\hss$\mathsurround=0pt#1$\hss}}}

\makeatother
\def\mathclap#1{\text{\hbox to 0pt{\hss$\mathsurround=0pt#1$\hss}}}

\usepackage[font=footnotesize]{caption}
\usepackage[font=footnotesize]{subcaption} 

\usepackage{relsize}
\usepackage{bm}
\usepackage{soul}
\usepackage{empheq}
\usepackage{placeins}

\begin{document}

\title{Accurate Indoor Radio Frequency Imaging using a New Extended Rytov Approximation for Lossy Media}

\author{Amartansh~Dubey, Samruddhi~Deshmukh,~\IEEEmembership{Graduate Student Members,~IEEE}, Li~Pan,~\IEEEmembership{Member,~IEEE}, Xudong~Chen,~\IEEEmembership{Fellow,~IEEE} and~Ross~Murch,~\IEEEmembership{Fellow,~IEEE}\thanks{This work was supported by the Hong Kong Research Grants Council with
		the General Research Fund grant 16211618 and the Collaborative Research Fund C6012-20G.}\thanks{A. Dubey, S. Deshmukh are with the Department of Electronic and Computer Engineering,
		Hong Kong University of Science and Technology (HKUST), Hong Kong, (e-mail:
		\protect{}{adubey@connect.ust.hk}).}\thanks{L. Pan is with Department of Sensor Physics, Halliburton, Singapore.}
		\thanks{X. Chen is with the Department of Electrical and Computer Engineering, National University of Singapore, Singapore.} \thanks{R. Murch is with the Department of Electronic and Computer Engineering
		and the Institute of Advanced Study both at the Hong Kong University
		of Science and Technology (HKUST), Hong Kong.}}

\maketitle


\begin{abstract}
Imaging objects with high relative permittivity and large electrical size remains a challenging problem in the field of inverse scattering. In this work we present a phaseless inverse scattering method that can accurately image and reconstruct objects even with these attributes. The reconstruction accuracy obtained under these conditions has not been achieved previously and can therefore open up the area to technologically important applications such as indoor Radio Frequency (RF) and microwave imaging. The novelty of the approach is that it utilizes a high frequency approximation for waves passing through lossy media to provide corrections to the conventional Rytov approximation (RA). We refer to this technique as the Extended Phaseless Rytov Approximation for Low Loss Media (xPRA-LM). Simulation as well as experimental results are provided for indoor RF imaging using phaseless measurements from 2.4 GHz based WiFi nodes. We demonstrate that the approach provides accurate reconstruction of an object  up to relative permittivities of $15+j1.5$ for object sizes greater than $20 \lambda$ ($\lambda$ is wavelength inside object). Even at higher relative permittivities of up to $\epsilon_r=77+j 7$, object shape reconstruction remains accurate, however the reconstruction amplitude is less accurate. These results have not been obtained before and can be utilized to achieve the potential of RF and microwave imaging in applications such as indoor RF imaging.   
\end{abstract}


%
\IEEEpeerreviewmaketitle

\section{Introduction}
\label{Sec_Intro}
Imaging objects with high complex permittivity $(\epsilon_r = \epsilon_R+ j \epsilon_I)$ and large size (sizes comparable to or greater than the free space wavelength $\lambda_0$) remains a challenging problem in the field of inverse scattering. The best existing linear and non-linear approaches have been shown to provide good reconstructions only when $\epsilon_R < 3$ and the scatterer is smaller than $\lambda_0$ \cite{chen2018computational, murch1990inverse}. Recent non-linear and deep learning based approaches \cite{chen2010, Xudongchen, chen2018computational, chen2020review} can moderately extend this range, but they either require high precision measurements in a controlled environment (such as an anechoic chamber) or only provide numerical results without experimental demonstration. Objects or scatterers with large electrical size and high complex refractive index or equivalently high complex relative permittivity often violate the assumptions upon which current inverse scattering techniques are based. Due to this, despite the plethora of existing theoretical techniques, there is limited practical use of inverse scattering methods in applications such as RF indoor imaging \cite{chen2010, chen2018computational, depatla2015x, chen2020review}, non-destructive evaluation/fault imaging (in electrical and mechanical structures) \cite{DubeyTxline, jing2018approximate, NC1} and other microwave imaging applications \cite{jing2018, pastorino2010microwave, NE1, benny2020overview, NE2, nature1} including microwave medical imaging \cite{benny2020overview}.

To overcome limitations associated with formal inverse scattering techniques, ad-hoc methods have been proposed. For example, in the area of indoor RF imaging, the use of radio tomographic imaging (RTI) methods, where a large body of research now exists, has been developed \cite{Patwari2010, Patwari2014, Patwari2015, Patwari2017, 1Patwari2013}. RTI utilizes a straightforward straight ray reconstruction approach that is not formally justified by the underlying wave theory. Nevertheless reasonably accurate localization and low resolution shape reconstruction is achieved for moving objects \cite{Patwari2010, Patwari2014, Patwari2015, Patwari2017, 1Patwari2013}. Due to the ad-hoc nature of the approach, generally only experimental results are provided without analysis of the validity range or inherent accuracy of the approach \cite{sood2020demonstrating}. Furthermore, estimates of the permittivity of the scatterers is also not achieved and the reconstructions focus only on the presence or absence of an object. Due to the success of the RTI approach, attempts have also been made at combining it with tools from inverse scattering \cite{depatla2015x, Mostofi2017, Savazzi2014, Dubey2021}. Even with these enhancements only one technique \cite{Dubey2021} attempts to reconstruct permittivity but this still does not provide any theoretical or numerical analysis for the range of validity and also relies on experiments. Nevertheless the success of these ad-hoc techniques suggests that accurate straightforward formal inverse scattering techniques are potentially possible. 

In this paper, we propose a phaseless inverse scattering method which can handle objects with large electrical size and large complex permittivity while using a practical measurement system. We demonstrate the technique for the specific use-case of indoor imaging using phaseless signals (we use WiFi received signal strength indicator (RSSI)) where the domain of interest (DOI) can be as large as $20$  to $100 \times \lambda_0$ for rooms and the relative permittivity of scatterers within the indoor environment can range from $\epsilon_R =$ 2 to 77 ($\epsilon_R=77$ for water and some parts of the human body \cite{4562803, Productnote, ahmad2014partially}). The method could be applied to a wider range of use-cases where scatterers have high complex permittivity and are electrically large \cite{DubeyTxline, jing2018approximate, depatla2015x, benny2020overview}.


\subsection{Motivation and Contributions}
\label{Sec_contri}
The motivation behind this work is to formulate a phaseless inverse scattering method which can handle lossy scatterers with  high permittivity and large electrical size (comparable or larger than free space wavelength $\lambda_0$). The focus is on formal inverse scattering methods that utilize straightforward phaseless measurements so that synchronization is not required between measurement nodes. To achieve this, we derive corrections to the Rytov Approximation (RA) by incorporating a high frequency approximation to the waves in lossy media. The resulting extended Phaseless Rytov Approximation for low loss media (xPRA-LM) can provide accurate reconstruction of shape but more importantly accurate reconstruction of the imaginary part of the contrast function (which is a function of both $\epsilon_R$ and $\epsilon_I$ and can be used for material identification as we shall show later). While it cannot accurately estimate the real part of the contrast function, the accurate reconstruction of the imaginary component of the contrast function is obtained even under very high complex permittivity and large electrical size scenarios. The technique is therefore very useful in applications such as indoor imaging as it provides high resolution shape estimation as well as being able to differentiate between materials and objects within the indoor environment. 

To summarize, the key contributions of this work are:
\begin{enumerate}
	\item We derive a new phaseless inverse scattering method denoted as xPRA-LM for low-loss media. It involves making use of a high frequency approximation in lossy media and provides corrections to RA by approximately estimating the gradient of the scattered field inside the scatterer (which is neglected in RA).
	\item Along with corrections to RA, we also extend it to a background subtraction framework where change in RSSI is linearly related to change in DOI contrast profile. This is very useful for applications such as indoor imaging and allows the removal of distortions due to scattering and reflections from the stationary background (ceiling, walls and floor) as well as clutter.
	\item Our corrections to RA are also used to provide explanations for two crucial previously unexplained/unexplored characteristics of RA including the validity of RA in lossy media and ``crosstalk" between the real and imaginary parts of the contrast function reconstruction \cite{Murch1996}. 
	\item We also show by simulation and experiment that xPRA-LM provides accurate shape reconstruction as well as providing estimates of the imaginary part of the contrast function. The validity range of xPRA-LM only requires objects to be of low loss ($\epsilon_R \gg \epsilon_I$) while $\epsilon_R$ can be arbitrarily large. Results are provided for scatterers  having a large range in complex permittivity from $1.1+j0.1$ to $77+j7$ and large electrical sizes up to $5  \lambda_0$.

\end{enumerate}


To the best of our knowledge, the xPRA-LM model derived in this work provides the first analysis of inverse scattering problems where there is no strict limit on relative permittivity values or scatterer size. Also, the background subtraction framework incorporated in xPRA-LM is extremely useful for practical applications such as indoor imaging and to the best of our knowledge, none of the existing phaseless inverse scattering methods \cite{klibanov1992phaseless, Xudongchen, chen2010subspace, murch1988newton} can be formulated to handle background subtraction.

Overall, the accuracy of our formulation in imaging lossy objects can be extremely useful for applications such as indoor imaging (using WiFi) where most objects satisfy a low-loss, high frequency assumption. For example, in indoor imaging almost all objects have small loss tangent $\delta = \epsilon_I/\epsilon_R \ll 1$ at 2.4 GHz. The low loss assumption utilized in xPRA-LM allows the real part of permittivity to take arbitrarily large values while keeping $\delta \ll 1$ (for example, pure water and some parts of the human body have very high permittivity $\epsilon_r= 77 + j 7$ with $\delta \approx 0.1$). This contrast with previous work which requires permittivity  $\epsilon_R$ to be small ($\epsilon_R< 2$ or $3$).

Organization of paper: Section \ref{Sec_Prelim} provides preliminary background for the concepts of complex refractive index, complex permittivity and the formulation of RA in phaseless form. The derivation of the proposed xPRA-LM technique is described in section \ref{Sec_RytovOptics}. Simulation and experimental results are provided in Section \ref{Sec_Results} and \ref{Sec_exp} to demonstrate the very good performance of the technique. These show that accurate RF imaging can be performed using xPRA-LM.

\section{Preliminaries} 
\label{Sec_Prelim}
\subsection{Low-Loss Dielectric Medium}

In lossy media, the refractive index and relative permittivity become complex and can be represented respectively as \cite{griffiths2005},
\begin{equation}
	\label{Eq_complexNu}
	\begin{aligned}
		\nu &= \nu_R + j \nu_I\\
		\epsilon_r &= \epsilon_R + j \epsilon_I
	\end{aligned}
\end{equation}
We consider monochromatic waves in this work and hence, for brevity, we do not include the dependence on frequency explicitly when writing $\nu$ and $\epsilon_r$. The relation between refractive index and relative permittivity is given by $\nu^2 = \epsilon_r$ \cite{yang2009effective, groth2016numerical, zhang2020generalized, zhang2015refractive, chang2005ray, yang1995light, jones1970ray}, hence,
\begin{equation}
	\label{Eq_complexNu1}
	\begin{aligned}
		(\nu_R + j \nu_I)^2 = \epsilon_R + j \epsilon_I
	\end{aligned}
\end{equation}
Equating real and imaginary parts gives,
\begin{equation}
	\label{Eq_complexNu3}
	\begin{aligned}
		\nu_R^2 - \nu_I^2 = \epsilon_R , \quad
		\nu_R \nu_I = \epsilon_I/2
	\end{aligned}
\end{equation}
and these equations can be solved to express $\nu_R$ and $\nu_I$ as,
\begin{subequations}
	\label{Eq_complexNu_R}
	\begin{align}
		\nu_R &= \bigg\{\frac{1}{2} \bigg(\sqrt{\epsilon_R^2+\epsilon_I^2} + \epsilon_R\bigg) \bigg\}^{1/2} \\
		\nu_I &= \bigg\{\frac{1}{2} \bigg(\sqrt{\epsilon_R^2+\epsilon_I^2} - \epsilon_R\bigg) \bigg\}^{1/2} 
	\end{align}
\end{subequations}
A medium can be characterized as low loss if  $\epsilon_I \ll \epsilon_R$. More specifically, the loss tangent for the medium is often defined as
\begin{equation}
 \delta = \epsilon_I/\epsilon_R
\end{equation} 
 so that low loss can also be expressed as $\delta \ll 1$.  Therefore, for low-loss media, (\ref{Eq_complexNu_R}a) can be simplified using the binomial expansion as,
\begin{equation}
	\label{Eq_complexNu_R_LL}
	\begin{aligned}
		\nu_R &= \bigg\{\frac{1}{2} \bigg(\sqrt{\epsilon_R^2+\epsilon_I^2} + \epsilon_R\bigg) \bigg\}^{1/2} \\
		&=\bigg\{\frac{1}{2} \bigg(\epsilon_R \sqrt{1+\delta^2} + \epsilon_R\bigg) \bigg\}^{1/2} \\
		&\approx\sqrt{\epsilon_R + {\frac{1}{4}  \delta^2 \epsilon_R} } \ \ \ \  \text{expand $\sqrt{1+\delta^2} \approx 1+\frac{1}{2}\delta^2$}\\
	\end{aligned}
\end{equation}
Similarly, (\ref{Eq_complexNu_R}b) can be approximated as
\begin{equation}
	\label{Eq_complexNu_I_LL}
	\begin{aligned}
		\nu_I  = \frac{\epsilon_I}{2 \sqrt{\epsilon_R} } = \frac{1}{2} \delta \sqrt{\epsilon_R}
	\end{aligned}
\end{equation}
Under practical constraints, (\ref{Eq_complexNu_R_LL}) can be further simplified as $\nu_R = \sqrt{\epsilon_R}$. 

\subsection{Phaseless Rytov Approximation}
\label{Sec_Rytov}
As background to the formulation of phaseless RA, let $E_i(\bm{r})$ be the incident field at any point inside DOI in the absence of a scatterer so that it is a solution of the free space Helmholtz wave equation,
\begin{equation}
\label{Eq_Hzfree}
\begin{aligned}
(\nabla^2 + k_0^2) E_i(\bm{r}) = 0
\end{aligned}
\end{equation}
where $k_0=2\pi/\lambda_0$ is the free space wavenumber. In the presence of scatterers, the total field is written $E(\bm{r})$ which is a solution of the inhomogeneous Helmholtz wave equation, 
\begin{equation}
\label{Eq_Hz}
\begin{aligned}
(\nabla^2 + k_0^2 \nu^2(\bm{r})) E(\bm{r}) = 0
\end{aligned}
\end{equation}
RA is then formulated by introducing the complex wavefront function  $\phi_s(\bm{r})$ for the scattered field as
\begin{equation}
\label{Eq_rytTfield}
\begin{aligned}
\frac{E(\bm{r})}{E_i(\bm{r})} &= e^{jk\phi_s(\bm{r}) }
\end{aligned}
\end{equation}

%

Substituting (\ref{Eq_rytTfield}) in (\ref{Eq_Hz}) and using (\ref{Eq_Hzfree}) provides the non-linear differential equation \cite{wu2003wave},
\begin{subequations}
	\label{Eq_rytdiffeq3}
	\begin{align}
		\nabla^2\tilde{E}& (\bm{r}) + k_0^2 \tilde{E}(\bm{r})= \nonumber \\ &-k^2[(\nu(\bm{r})^2-1)-\nabla \phi_s(\bm{r}) \cdot \nabla \phi_s(\bm{r})] E_i(\bm{r})
	\end{align}
	\begin{align}
	\tilde{E}(\bm{r}) = E_i(\bm{r}) \ln\biggl[\frac{E(\bm{r})}{E_i(\bm{r})}\biggr]
\end{align}
\end{subequations}
which can be written in the form of an integral solution as,
\begin{equation}
	\label{Eq_rytov1}
	\begin{aligned}
		\frac{E(\bm{r})}{E_i(\bm{r})}  = \exp\bigg(\frac{k^2}{E_i(\bm{r})} \int_{A} & g(\bm{r}, \bm{r'})  \bigg[\nu(\bm{r'})^2-1 -  \\ & \nabla \phi_s(\bm{r'}) \cdot \nabla \phi_s(\bm{r'})\bigg] E_i(\bm{r'}) d\bm{r'}^2\bigg)
	\end{aligned}
\end{equation}

Multiplying (\ref{Eq_rytov1}) by its conjugate and taking $\log_{10}$ both sides provides the phaseless form of (\ref{Eq_rytov1}) in terms of the total and incident power (in dB), 
\begin{equation}
	\label{Eq_RytovInt}
	\begin{aligned}
		P(\bm{r}) & [\text{dB}] = P_i(\bm{r})[\text{dB}]  \ +  \\ & C_0 \cdot \operatorname{Re}\bigg(\frac{k^2}{E_i(\bm{r})}  \int_{A} g(\bm{r}, \bm{r'})  \chi_{\text{RI}}(\bm{r'}) E_i(\bm{r'}) d\bm{r'}^2\bigg) 
	\end{aligned}
\end{equation}
where, $\operatorname{Re}$ denotes real part operator, $C_0 = 20 \log_{10} e$ is a constant and,
\begin{equation}
	\label{Eq_RytovInt_CF}
	\begin{aligned}
	\chi_{\text{RI}}(\bm{r'}) & = \nu(\bm{r'})^2-1 - \nabla \phi_s(\bm{r'}) \cdot \nabla \phi_s(\bm{r'}) .
	\end{aligned}
\end{equation}
$\chi_{\text{RI}}(\bm{r})$ is referred to in the remainder of this paper as the contrast function for the Rytov Integral (RI) (\ref{Eq_RytovInt_CF}).

Solving (\ref{Eq_RytovInt}) as an inverse problem requires us to find both $\nu(\bm{r'}), \phi_s(\bm{r'})$ inside the scatterer from knowledge of only $P(\bm{r}) - P_i(\bm{r})$ at the measurement boundary. This is a difficult non-linear ill-posed problem. To simplify this, RA, neglects the term $\nabla \phi_s \cdot \nabla \phi_s$ which is generally considered appropriate under weak scattering ($\epsilon_r \approx 1$). This results in the phaseless form of RA,
\begin{subequations}
	\label{Eq_RytovApp}
	\begin{align}
		P(\bm{r}) & [\text{dB}] = P_i(\bm{r})[\text{dB}]  \ + \nonumber \\ & C_0 \cdot \operatorname{Re}\bigg(\frac{k^2}{E_i(\bm{r})}  \int_{A} g(\bm{r}, \bm{r'})  \chi_{\text{RA}}(\bm{r'}) E_i(\bm{r'}) d\bm{r'}^2\bigg)
	\end{align}
	\begin{align}
		\chi_{\text{RA}}(\bm{r}') = \nu(\bm{r}')^2-1 = \epsilon_r-1
	\end{align}
\end{subequations}
The difference between (\ref{Eq_RytovInt}) and (\ref{Eq_RytovApp}) is the omission of $(\nabla \phi_s\cdot \nabla \phi_s)$ in order to make it linear in $P(\bm{r})-P_i(\bm{r})$ and the unknown permittivity $\epsilon_r(\bm{r})$. However, this makes RA useful only for weak scattering with $\epsilon_R \approx 1$, hence making it futile for  applications such as RF indoor and microwave imaging.

In the remainder of this paper, we formally refer to RA as denoting (\ref{Eq_RytovApp}) and the Rytov Integral (RI) as denoting (\ref{Eq_RytovInt}) and use the terms to distinguish between when $\nabla \phi_s\cdot \nabla \phi_s$ is or is not neglected. The contrast functions of RA and RI are consequently denoted as $\chi_{\text{RA}}$ (\ref{Eq_RytovApp}b) and $\chi_{\text{RI}}$ (\ref{Eq_RytovInt_CF}) respectively.


\section{The Extended Phaseless Rytov Approximation Formulation for Low Loss Media}
\label{Sec_RytovOptics}

Conventional RA is valid under the weak scattering assumption, where the term $\nabla \phi_s \cdot \nabla \phi_s$ is small and can be neglected from the contrast function $\chi_{\text{RA}}$. By doing so, as shown in (\ref{Eq_RytovApp}), the real part of the reconstruction $\operatorname*{Re}(\chi_{\text{RA}})$ is physically interpreted as \textit{phase contrast} which linearly relates to the real part of permittivity $(\epsilon_R)$ and the imaginary part of the reconstruction $\operatorname*{Im}(\chi_{\text{RA}})$ is interpreted as an \textit{attenuation or loss contrast} which linearly relates to the imaginary part of permittivity $(\epsilon_I)$. 

For large scatterers with high permittivity, the term $\nabla \phi_s \cdot \nabla \phi_s$ is significant and cannot be neglected. Estimation of $\nabla \phi_s \cdot \nabla \phi_s$ is difficult as it requires solving the intractable non-linear equation (\ref{Eq_rytov1}) during the inversion process and to the best of our knowledge, has not been performed previously \cite{enright1992towards}. 

In the next section, we use a high frequency approximation to the field in low-loss media to approximately find the gradient of the scattered field $\nabla \phi_s$. Using this, we can approximate $\chi_{\text{RI}}$ and hence use RI for imaging under strong scattering and high frequency conditions. We show that due to the presence of the term $\nabla \phi_s \cdot \nabla \phi_s$ in $\chi_{\text{RI}}$, the real and imaginary parts of the reconstruction are no longer linearly related to $\epsilon_R$ and $\epsilon_I$ (unlike RA) which completely changes the interpretation of reconstructions provided by RA. Combining this derived expression for RI with a background subtraction framework provides our proposed method, xPRA-LM. 

\subsection{Rays in Lossy Media}
\label{Sec_rayoptics}
When the size of the scatterer is larger than the incident wavelength $\lambda_0$, we can use concepts such as rays, ray optics or geometrical optics (GO), to describe wave scattering. Conventional ray optics usually ignores the absorbing effects of loss in the medium \cite{bornwolf, jones1970ray, murch1994evaluation}. Despite ray based techniques being a decades old approach, there are surprisingly limited ray formulations for lossy media. When media is lossy, the refractive index becomes complex and the waves inside the media become inhomogeneous  \cite{yang2009effective, groth2016numerical, zhang2020generalized, zhang2015refractive, chang2005ray, yang1995light, jones1970ray}. Inhomogeneous waves exhibit the property that the \textit{planes of constant phase} are no longer parallel to the \textit{planes of constant amplitude} \cite{yang2009effective, groth2016numerical, zhang2020generalized, zhang2015refractive, chang2005ray, yang1995light, jones1970ray}. In the remainder of this subsection, we deal with homogeneous plane waves (HPW) in lossless media that is incident on lossy media and become inhomogeneous plane waves (IPW) inside the lossy media. 

\begin{figure}[!h]
	\centering
	\includegraphics[width=2.5in]{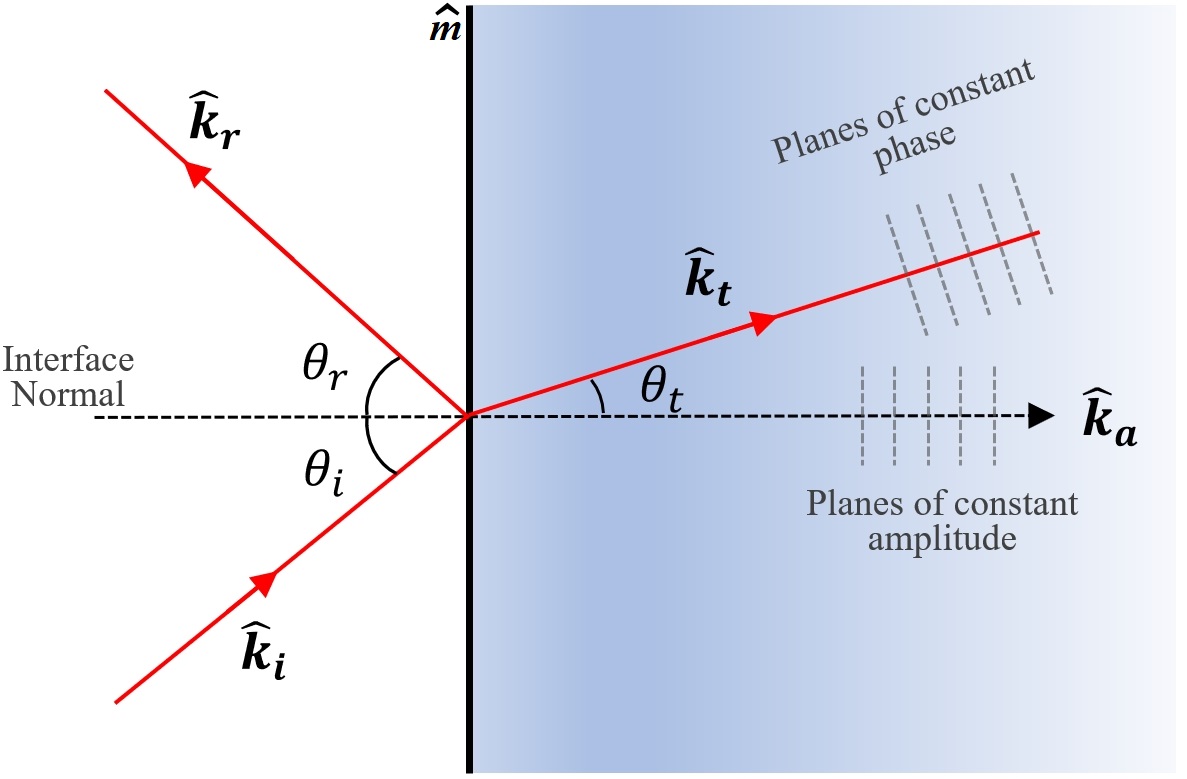}
	\caption{Free space to lossy media interface. The homogeneous plane wave (HPW) in lossless media becomes inhomogeneous plane wave (IPW) inside the lossy media (with refractive index $\nu = \nu_R+ j \nu_I$).}
	\label{interface}
	\vspace{-0.3\baselineskip}
\end{figure}

Fig. \ref{interface} illustrates a vacuum/air to lossy dielectric interface. The wave equations in free space (on left) and lossy media (on right, with complex refractive index, $\nu_R+j\nu_I$) can be given as (\ref{Eq_Hzfree}) and (\ref{Eq_Hz}) respectively. 

The incident field is partially reflected and transmitted at the interface. Using Snell's law $(\sin\theta_i = (\nu_R + j\nu_I)\sin\theta_t)$, we find that the angle of refraction becomes a complex quantity, which cannot be interpreted using conventional GO. To make the analysis geometrically more intuitive and remove complex angles, the concept of effective (or apparent) refractive index has been introduced \cite{yang1995light, yang2009effective, zhang2020generalized} and is also used in our work. 

The idea of effective refractive index is to decompose the mathematical form of IPW by expressing it as a linear combination of vectors normal to the constant phase and amplitude planes and then invoke Snell's law separately for the refraction and attenuation components of IPW (see \cite{chang2005ray, zhang2015refractive, zhang2020generalized, groth2016numerical, yang1995light, yang2009effective}). Using this concept, the wave-vectors of the incident, reflected and IPW transmitted field can be given as,
\begin{equation}
\label{Eq_complexwavevector}
\begin{aligned}
\bm{k}_i = k_0 \bm{\hat{k}_i}, \ \ \
\bm{k}_r = k_0 \bm{\hat{k}_r},\ \ \
\bm{k}_t  = k_0 (V_R \bm{\hat{k}_t} + j V_I \bm{\hat{k}_a})
\end{aligned}
\end{equation}
where, $V_R$ and $V_I$ are termed the effective real and imaginary parts of the refractive index respectively \cite{chang2005ray, zhang2015refractive, zhang2020generalized, groth2016numerical, yang1995light, yang2009effective}. The unit vectors $\bm{\hat{k}_t}$ and $\bm{\hat{k}_a}$ are in the normal direction to the constant phase and amplitude planes respectively. Using wave vectors in (\ref{Eq_complexwavevector}), we can write the incident, reflected and transmitted fields as,
\begin{subequations}
	\label{Eq_complexwavevectorfield}
	\begin{align}
	E_i(\bm{r}) &= A_0 (\bm{r}) \exp{(jk_0 \ \bm{\hat{k}}_i \cdot \bm{r}  )} \\
	E_r(\bm{r}) &= A_r(\bm{r}) \exp{(jk_0 \ \bm{\hat{k}}_r \cdot \bm{r}  )} \\
	E_t(\bm{r}) &= A_t(\bm{r}) \exp{(jk_0 \ (V_R \bm{\hat{k}_t} \cdot \bm{r}  + jV_I \bm{\hat{k}_a} \cdot \bm{r}))}
	\end{align}
\end{subequations}
Snell's law can now be derived for this case. The phase of the incident, reflected and transmitted fields should match tangentially at the media interface. Letting the vector tangential to the interface be $\bm{\hat{m}}$ (see Fig. \ref{interface}) we get,
\begin{equation}
	\label{Eq_SnellLaw}
	\begin{aligned}
	\bm{\hat{m}} \cdot \bm{\hat{k}_i} = 	\bm{\hat{m}} \cdot \bm{\hat{k}_r} = \bm{\hat{m}} \cdot	(V_R \bm{\hat{k}_t} + j V_I \bm{\hat{k}_a} ) 
	\end{aligned}
\end{equation}
The wave vector for the incident and reflected waves should be real so that by  equating real parts in (\ref{Eq_SnellLaw}) gives,
\begin{equation}
	\label{Eq_SnellLaw1}
	\begin{aligned}
	\bm{\hat{m}} \cdot \bm{\hat{k}_i} = \bm{\hat{m}} \cdot \bm{\hat{k}_r} = \bm{\hat{m}} \cdot	(V_R \bm{\hat{k}_t} )\\
	\sin\theta_i = \sin\theta_r = V_R \sin\theta_t.
	\end{aligned}
\end{equation}
and is the equivalent of Snell's law. Similarly, equating imaginary parts in (\ref{Eq_SnellLaw}) gives,
\begin{equation}
\label{Eq_SnellLaw2}
\begin{aligned}
\bm{\hat{m}} \cdot	(V_I\bm{\hat{k}_a} ) = 0  \implies \bm{\hat{m}} \perp \bm{\hat{k}_a}
\end{aligned}
\end{equation}
This implies that the plane of constant amplitude (i.e. plane normal to $\bm{\hat{k}_a}$) is parallel to the interface as shown in Fig. \ref{interface}. 

On substituting IPW (\ref{Eq_complexwavevectorfield}c) into wave equation (\ref{Eq_Hz}) and using (\ref{Eq_SnellLaw1}), we arrive at the following relation between actual and effective refractive index, 
\begin{subequations}
	\label{Eq_effective}
	\begin{align}
	V_R^2 - V_I^2 = \nu_R^2 - \nu_I^2 \\
	V_R {V_I}\cos\theta_t 	= \nu_R \nu_I  && \text{where, $\bm{\hat{k}_t} \cdot \bm{\hat{k}_a}= \cos\theta_t$}
	\end{align}
\end{subequations}
Equation (\ref{Eq_effective}) can be solved by substituting ${V_I}$ from (\ref{Eq_effective}b) into (\ref{Eq_effective}a) and then solving the quartic equation to obtain $V_R$ as, 
\begin{equation}
	\label{Eq_effectivefinal}
	\begin{aligned}
	V_R &= \bigg\{\frac{1}{2} \bigg(\sqrt{(\nu_R^2 - \nu_I^2)^2 +  4 \bigg[\frac{ \nu_R \nu_I}{\cos\theta_t}\bigg]^2} + \nu_R^2-\nu_I^2  \bigg) \bigg\}^{1/2} 
	\end{aligned}
\end{equation}
The value of $V_I$ can be estimated using (\ref{Eq_effectivefinal}) and (\ref{Eq_effective}b). 

By imposing the low-loss assumption $(\epsilon_I\ll \epsilon_R)$ and using (\ref{Eq_SnellLaw1}) to express $\cos\theta_t$ in terms of $\sin\theta_i$ in (\ref{Eq_effectivefinal}), we can use the binomial expansion \cite{chang2005ray} to approximate $V_R$, $V_I$ and $\theta_t$ as (and use (\ref{Eq_complexNu_R_LL}) and (\ref{Eq_complexNu_I_LL}) to express in terms of permittivity), 
\begin{equation}
	\label{Eq_VRVIlowloss1}
	\begin{aligned}
		V_R & \approx \nu_R \bigg(1+\frac{\sin^2\theta_i}{2(\nu_R^2 - \sin^2\theta_i)} \delta^2 \bigg) \\
		&  \approx \nu_R  \approx \sqrt{\epsilon_R} 
	\end{aligned}
\end{equation}
\begin{equation}
	\label{Eq_VRVIlowloss2}
	\begin{aligned}
		V_I &\approx \frac{\nu_R \nu_I}{\sqrt{\nu_R^2 - \sin^2\theta_i}}\bigg(1-\frac{\nu_R^2 \sin^2\theta_i}{2(\nu_R^2-\sin^2\theta_i)}\delta^2 \bigg) \\
		& \approx \frac{\nu_R \nu_I}{\sqrt{\nu_R^2 - \sin^2\theta_i}} \approx \frac{\epsilon_I}{2 \sqrt{\epsilon_R - \sin^2\theta_i}}
	\end{aligned}
\end{equation}


Using the previous relations, we can find the expression for the ray in Fig. \ref{interface} along ray path $\bm{{dr}} =dr \ \bm{\hat{k}_t}$ as,
\begin{equation}
	\label{Eq_raytraced}
	\begin{aligned}
		E_t(\bm{r}) &= A_t \exp{(jk_0 \ (V_R \bm{\hat{k}_t} \cdot \bm{{dr}}  + jV_I \bm{\hat{k}_a} \cdot \bm{{dl}} ) )} \\
		& = A_t \exp{(jk_0 \ (V_R dr \bm{\hat{k}_t} \cdot \bm{\hat{k}_t}  + jV_I dr \bm{\hat{k}_a} \cdot \bm{\hat{k}_t} ) )}\\
		& = A_t \underbrace{\exp{(-k_0 \ V_I dr \cos\theta_t  )}}_\text{attenuation term}\exp{(jk_0 \ V_R dr )} 
	\end{aligned}
\end{equation}
where, $A_t, V_R, V_I$ are functions of $\bm{r}$. The ray equation for IPW (\ref{Eq_raytraced}) is for a ray passing through differential element $dr$ along ray path $\bm{{dr}} =dr \ \bm{\hat{k}_t}$. For an extended area of a lossy scatterer with piece-wise homogeneous distribution of refractive index, (\ref{Eq_raytraced}) can be written in terms of path integral along the ray direction as, 
\begin{equation}
\label{Eq_raytraced1}
\begin{aligned}
E_t(\bm{r}) & =A_t \ {\exp{\biggl(-k_0 \int\displaylimits_{\mathclap{\text{along $\bm{\hat{k}_t}$}}} V_I \bm{\hat{k}_a} \bm{dr} \biggr)}}\exp{\biggl(jk_0 \int\displaylimits_{\mathclap{\text{along $\bm{\hat{k}_t}$}}} \ V_R \bm{\hat{k}_t} \bm{dr} \biggr)}\\
& =A_t \ {\exp{\biggl(-k_0 \int\displaylimits_{\mathclap{\text{along $\bm{\hat{k}_t}$}}} V_I \cos\theta_t  dr \biggr)}}\exp{\biggl(jk_0 \int\displaylimits_{\mathclap{\text{along $\bm{\hat{k}_t}$}}} \ V_R dr \biggr)} 
\end{aligned}
\end{equation}

In the next section, we use this form to approximate the term $\nabla \phi_s\cdot \nabla \phi_s$ in order to enhance the accuracy of RA. Note that (\ref{Eq_complexwavevectorfield}) represents the first order ray inside a piece-wise homogeneous scatterer. There will be higher order rays inside the scatterer due to multiple scattering. However, for a lossy scatterer, it is shown that the first order ray is a good approximation \cite{yang2009effective} as higher order rays will contain low energy (see Appendix \ref{MultScattering} for details).

\subsection{RI for Low Loss Media at High Frequencies}
\label{Sec_rayRytov}
To approximate the term $\nabla \phi_s \cdot \nabla \phi_s$ in RI, we start by equating the total field inside the scatterer  (\ref{Eq_rytTfield}) using the ray equation (\ref{Eq_raytraced1}) to obtain
\begin{equation}
\label{Eq_rytHFtotal}
\begin{aligned}
& E_i  (\bm{r})  e^{jk_0\phi_s(\bm{r})} = \\ & A_t(\bm{r}) {\exp{\biggl(-k_0 \int V_I(\bm{r}) \cos\theta_t \  dr \biggr)}} \exp{\biggl(jk_0 \int V_R(\bm{r})\  d{r}\biggr) }
\end{aligned}
\end{equation}
Substituting the incident field from (\ref{Eq_complexwavevectorfield}a) as  $E_i(\bm{r}) = A_0(\bm{r})  e^{j k_0 \bm{\hat{k}_i \cdot r}}$ gives,
\begin{equation}
\label{Eq_rytHFtotal1}
\begin{aligned}
& \phi_s(\bm{r})\\ & = \frac{1}{jk_0 }\text{ln}\biggl[\frac{A_t}{E_i }  {\exp{\biggl(-k_0 \int V_I \cos\theta_t dr \biggr)}} \exp{\biggl(jk_0 \int V_R\  dr\biggr) }\biggr]\\
& =  \frac{1}{jk_0} \text{ln}\biggl[\frac{A_t}{A_0}\biggr] + \biggl[\int V_R\  dr - \bm{\hat{k}_i \cdot r} +j\int V_I \cos\theta_t dr \biggr]
\end{aligned}
\end{equation}
Note that the quantities $A_t, A_0, V_R, V_I, E_i$ are functions of $\bm{r}$ in (\ref{Eq_rytHFtotal1}) and for brevity we do not show this dependence in the remainder of this paper. Taking the gradient of (\ref{Eq_rytHFtotal1}) gives (recall from (\ref{Eq_raytraced1}), $\bm{{dr}} =dr \ \bm{\hat{k}_t}$),
\begin{equation}
\label{Eq_rytHFphasegrad}
\begin{aligned}
 \nabla \phi_s & (\bm{r})= \\ & \frac{-j}{k_0}\biggl[ \nabla \underbrace{   \text{ln}\biggl(\frac{A_t}{A_0}\biggr)}_{\text{replaced as } \tilde{A}} \biggr] + \biggl[\big(V_R\ \bm{\hat{k}_t} - \bm{\hat{k}_i}\big) + j V_I \bm{\hat{k}_a} \biggr]
\end{aligned}
\end{equation}
so that
\begin{equation}
\label{Eq_rytHFphasegrad2}
\begin{aligned}
 \nabla & \phi_s  (\bm{r}) \cdot \nabla \phi_s(\bm{r}) = \\
& \biggl[V_R^2 + 1 -2 V_R (\bm{\hat{k}_t} \cdot \bm{\hat{k}_i}) - V_I^2 + 2 j (V_R \bm{\hat{k}_t}-\bm{\hat{k}_i})  \cdot V_I \bm{\hat{k}_a} \biggr] \\ & -  \frac{1}{k_0^2} (\nabla \tilde{A} \cdot \nabla \tilde{A} ) 
  -j\frac{2}{k_0} (\nabla \tilde{A}) \biggl[\big(V_R\ \bm{\hat{k}_t} - \bm{\hat{k}_i}\big) +  j V_I \bm{\hat{k}_a} \biggr]
\end{aligned}
\end{equation}
We note from Fig. \ref{interface}, $\bm{\hat{k}_i} \cdot \bm{\hat{k}_a} = \cos\theta_i$ and $\bm{\hat{k}_t} \cdot \bm{\hat{k}_i} = \cos\theta_{s}$ where $\theta_{s}$ is the scattering angle. We can now write (\ref{Eq_rytHFphasegrad2}) as
\begin{equation}
\label{Eq_rytHFphasegrad22}
\begin{aligned}
& \nabla  \phi_s(\bm{r}) \cdot \nabla \phi_s(\bm{r})  = \\
& \biggl[V_R^2 + 1 -2 V_R \cos\theta_{s} - V_I^2 + 2 j (V_R V_I \cos\theta_t - V_I \cos\theta_i)  \biggr] \\
&  - \frac{1}{k_0^2} (\nabla \tilde{A} \cdot \nabla \tilde{A} )  -j\frac{2}{k_0} (\nabla \tilde{A})\big(V_R\ \bm{\hat{k}_t} - \bm{\hat{k}_i}\big)  +   \frac{2}{k_0} (\nabla \tilde{A}) V_I \bm{\hat{k}_a}
\end{aligned}
\end{equation}
Separating out real and imaginary terms gives,
\begin{equation}
\label{Eq_rytHFphasegrad3}
\begin{aligned}
& \nabla \phi_s(\bm{r}) \cdot \nabla \phi_s(\bm{r}) = \\
& \biggl[V_R^2 + 1 -2 V_R \cos\theta_{s} - V_I^2 - \frac{1}{k_0^2} (\nabla \tilde{A} \cdot \nabla \tilde{A} ) + \\ 
&  \qquad \qquad \qquad \qquad  \qquad \qquad \qquad \qquad \qquad \frac{2}{k_0} (\nabla \tilde{A}) V_I \bm{\hat{k}_a} \biggr]  \\
& + 2j \biggl[ (V_R V_I \cos\theta_t - V_I \cos\theta_i) - \frac{1}{k_0} (\nabla \tilde{A})\big(V_R\ \bm{\hat{k}_t} - \bm{\hat{k}_i}\big) \biggr].
\end{aligned}
\end{equation}
Equation (\ref{Eq_rytHFphasegrad3}) provides an expression for $\nabla \phi_s \cdot \nabla \phi_s$ which is required in RI (\ref{Eq_RytovInt}) (neglected in RA (\ref{Eq_RytovApp})). Expanding the contrast function (\ref{Eq_RytovInt_CF}) of RI using (\ref{Eq_effective}) gives,
\begin{equation}
\label{Eq_rytovfulldB4}
\begin{aligned}
\chi_{\text{RI}}(\bm{r}) &= (\nu_R+j \nu_I)^2-1 - \nabla \phi_s \cdot \nabla \phi_s\\
& = \nu_R^2 -\nu_I^2 +2j \nu_R\nu_I-1 - \nabla \phi_s \cdot \nabla \phi_s\\
& = V_R^2 - V_I^2 + 2j V_R V_I \cos\theta_t -1 -\nabla \phi_s \cdot \nabla \phi_s
\end{aligned}
\end{equation}

Substituting $\nabla \phi_s \cdot \nabla \phi_s$ from (\ref{Eq_rytHFphasegrad3}) to (\ref{Eq_rytovfulldB4}) leads to cancellation of several terms and gives,
\begin{equation}
\label{Eq_rytovfulldB41}
\begin{aligned}
  \chi_{\text{RI}}&(\bm{r}) = \\
& {\bigg[2 V_R \cos\theta_{s} -2 + \frac{1}{k_0^2} (\nabla \tilde{A} \cdot \nabla \tilde{A} ) - \frac{2}{k_0} (\nabla \tilde{A}) V_I \bm{\hat{k}_a}  \bigg]} \\
& \qquad \qquad + j {\bigg[ 2 V_I \cos\theta_i + \frac{2}{k_0} (\nabla \tilde{A})\big(V_R(\bm{r})\ \bm{\hat{k}_t} - \bm{\hat{k}_i}\big) \bigg]}.
\end{aligned}
\end{equation}
Equation (\ref{Eq_rytovfulldB41}) can be further modified using (\ref{Eq_VRVIlowloss1}) and (\ref{Eq_VRVIlowloss2}) to replace $V_R$ and $V_I$ in terms of $\nu_R$ and $\nu_I$ under low-loss conditions as,
\begin{equation}
\label{Eq_rytovfulldB5}
\begin{aligned}
 \chi_{\text{RI}}&(\bm{r}) = \\
& {\bigg[2 (\nu_R \cos\theta_{s} -1) + \frac{1}{k_0^2} (\nabla \tilde{A} \cdot \nabla \tilde{A} ) - \frac{2}{k_0} (\nabla \tilde{A}) V_I \bm{\hat{k}_a}  \bigg]}  \\
& + j {\bigg[  2 \frac{\nu_R \nu_I}{\sqrt{\nu_R^2 - \sin^2\theta_i}} \cos\theta_i + \frac{2}{k_0} (\nabla \tilde{A})\big(\nu_R(\bm{r})\ \bm{\hat{k}_t} - \bm{\hat{k}_i}\big) \bigg] }
\end{aligned}
\end{equation}
Complex refractive index can also be expressed in terms of complex permittivity using (\ref{Eq_complexNu_R_LL}) and (\ref{Eq_complexNu_I_LL}). Using this, the final expression for the contrast in RI under low-loss, high frequency conditions is given by
	\begin{equation}
	\label{Eq_rytovfulldB6}
	\begin{aligned}
		\chi_{\text{RI}}&(\bm{r}) = \\ &\biggl(\underbrace{2 (\sqrt{\epsilon_R} \cos\theta_{s} -1)}_{\text{R}_1} + \underbrace{\frac{1}{k_0^2} (\nabla \tilde{A} \cdot \nabla \tilde{A} )}_{\text{R}_2 \text{ (crosstalk)}} - \underbrace{\frac{2}{k_0} (\nabla \tilde{A}) V_I \bm{\hat{k}_a}}_{\text{R}_3 \text{ (crosstalk)}}  \biggr)\\ 
		&+ j {\biggl( \underbrace{\frac{\epsilon_I}{\sqrt{\epsilon_R - \sin^2\theta_i}} \cos\theta_i}_{\text{I}_1} + \underbrace{\frac{2}{k_0} (\nabla \tilde{A})\big(\nu_R(\bm{r})\ \bm{\hat{k}_t} - \bm{\hat{k}_i}}_{\text{I}_2 \text{ (crosstalk)}}\big) \biggr) }
	\end{aligned} 
\end{equation}

To the best of our knowledge, this is the first time that the contrast function, $\operatorname*{Im}({\chi_{\text{RI}}})$, of RI has been given for lossy media. A significant aspect of (\ref{Eq_rytovfulldB6}) is that it can be clearly seen that both the real and imaginary parts of the contrast function contain significant terms. Furthermore, the imaginary part depends on both the real and imaginary parts of the permittivity and this describes the ``crosstalk" where even when the permittivity is real there will be a component in the imaginary part of the contrast function. Similarly, the real part of the contrast function also depends on both the real and imaginary parts of the permittivity.

Further simplification of (\ref{Eq_rytovfulldB6}) is possible. Because we are considering the high frequency regime (approximately $>1$ GHz) where $k_0$ is large, we can approximate (\ref{Eq_rytovfulldB6}) by ignoring the cross terms (R$_2$, R$_3$, I$_2$) as these include $1/k_0$ and $1/k_0^2$. This approximation will be valid as long as the spatial variation of the term $\tilde{A}=\ln (A_t/A_0)$ is small (so that $\nabla \tilde{A}$ is small). In applications such as indoor imaging, the objects considered are largely homogeneous and therefore the gradient of $\tilde{A}$ will be minimal inside and outside the objects. On the boundaries there will be a discontinuity and therefore our approximations will generally be accurate everywhere except at the boundaries of the objects where we can expect distortions in reconstruction amplitude. Based on these approximations, we can ignore the cross terms and rewrite (\ref{Eq_rytovfulldB6}) as,
\begin{equation}
	\label{Eq_rytovfulldB6-1}
	\begin{aligned}	
		\chi_{\text{RI}}(\bm{r}) &=2 (\sqrt{\epsilon_R} \cos\theta_{s} -1)
		+ j  \frac{\epsilon_I}{\sqrt{\epsilon_R - \sin^2\theta_i}} \cos\theta_i	
	\end{aligned} 
\end{equation}
From this equation we can see that there is a fundamental difference between the real and imaginary parts of the contrast function. In this form, the imaginary part $\operatorname*{Im}({\chi_{\text{RI}}})$ is a function of incident angle $\theta_i$ and both the real and imaginary parts of permittivity ($\epsilon_R, \epsilon_I$). While the real part of $\operatorname*{Re}({\chi_{\text{RI}}})$ is a function of the real part of the permittivity $\epsilon_R$ and the scattering angle $\theta_s$ which further depends on $\theta_i$ and $\epsilon_r$. Equation  (\ref{Eq_rytovfulldB6-1}) is also very different from previous RA formulations particularly in terms of $\operatorname*{Im}({\chi_{\text{RI}}})$ \cite{murch1990inverse, murch1994evaluation}.

The dependence of the imaginary part of the contrast function, $\operatorname*{Im}({\chi_{\text{RI}}})$ on $\theta_i$ instead of $\theta_s$ is the most important aspect of the derived result and is new to this field. The incident angle $\theta_i$ is only a function of the shape of the object. It does not change with the permittivity of the object. This implies that any distortion in the imaginary component, $\operatorname*{Im}({\chi_{\text{RI}}})$, of the reconstruction due to the presence of the $\theta_i$ terms is independent of the objects permittivity in our formulation (\ref{Eq_rytovfulldB6-1}). Therefore, if the imaginary component of the reconstruction of the contrast function is accurate at low permittivity, it is likely to be accurate for all permittivity levels. This is the reason we focus on this component in this paper.  

On the other hand, the real part of the reconstruction, $\operatorname*{Re}({\chi_{\text{RI}}})$  depends on $\theta_s$ which is dependent on $\theta_i$ as well as object permittivity. As demonstrated in a large body of previous literature, the real part has been shown to be accurate only for objects that have relative permitivities that deviate only slightly from unity \cite{wu2003wave, bates1976extended, feng2019}. In fact due to the dependency of $\operatorname*{Re}({\chi_{\text{RI}}})$ on $\theta_s$, the real part of the reconstruction for higher permittivity has significant distortion and is hence not very useful for imaging in applications such as indoor imaging \cite{wu2003wave, bates1976extended, feng2019}. 

We can also look at (\ref{Eq_rytovfulldB6-1}) from the perspective of Fermat's principle to gain more insight \cite{bates1976extended}. Under strong scattering  ($\epsilon_R\gg1$), and for the special cases of normal incidence ($\theta_i=0$) and forward scattering ($\theta_s=0$), our result  (\ref{Eq_rytovfulldB6-1}) reduces to refractive index $\chi_{\text{RI}} \approx 2 (\nu_R-1) + 2j \nu_I = 2(\nu-1)$. This agrees with Fermat's principle where the incremental phase change of a ray is directly related to the product of the path length along the ray and refractive index contrast $(\nu(\bm{r}) -1)$. In other words, the incremental phase change of a ray per wavelength should be proportional to $k_0(\nu-1)$. For conventional RA, it is known (using asymptotic techniques) that the incremental phase change per wavelength is $\frac{1}{2} k_0 (\nu^2 -1)$ which does not match the expected phase change as per Fermat's principal (\ref{Eq_rytovfulldB6-1}). Therefore xPRA-LM also appears to better satisfy the underlying physics of the problem. 

In the remainder of this work, we focus only on $\operatorname*{Im}({\chi_{\text{RI}}})$ since it is dependent only on $\theta_i$ and promises to provide accurate reconstruction. It has also not been explored previously.

\subsection{Final xPRA-LM form}
To make $\operatorname*{Im}({\chi_{\text{RI}}})$ in (\ref{Eq_rytovfulldB6-1}) tractable for imaging, it is necessary to remove the $\theta_i$ dependence in (\ref{Eq_rytovfulldB6-1}) so that $\operatorname*{Im}({\chi_{\text{RI}}})$ becomes only a function of permittivity ($\epsilon_r=\epsilon_R+ j \epsilon_I$). This can be performed by realizing that in the imaging process the object is often illuminated from a wide variety incidence directions (which is satisfied in the tomographic setup here) so that $\theta_i \in [-\pi/2, \pi/2]$. Therefore by taking an average over these angles, $\operatorname*{Im}({\chi_{\text{RI}}})$ can be written as
\begin{equation}
	\label{Eq_avgIm}
	\begin{aligned}
		\operatorname*{Im}({{\chi}_{\text{RI}}}) = \frac{1}{\pi} \int_{-\frac{\pi}{2}}^{\frac{\pi}{2}} \frac{\epsilon_I \cos\theta_i}{\sqrt{{\epsilon_R} - \sin^2\theta_i}} d\theta_i
	\end{aligned}
\end{equation}
This integral can be solved analytically (using substitution $u = \sin\theta_i$) to obtain
\begin{equation}
	\label{Eq_avgIm1}
	\begin{aligned}
		\operatorname*{Im}({{\chi}_{\text{RI}}})  = \frac{2 \epsilon_I}{\pi} \sin^{-1} \biggl(\frac{1}{\sqrt{\epsilon_R}}\biggr)
	\end{aligned}
\end{equation} 
Equation (\ref{Eq_avgIm1}) represents the final form of our imaging formulation and provides an expression for the imaginary part of the contrast function straightforwardly in terms of object permittivity ($\epsilon_I$ and $\epsilon_R$). Furthermore, with the contrast functions' analytical relation to $\epsilon_I$ and $\epsilon_R$, we will be able to distinguish between different types of materials in the DOI. 

To obtain further insight into  (\ref{Eq_avgIm1}) we can also consider the two limits of weak and strong scattering. For strong scattering, $\epsilon_R \gg 1, \delta \ll 1$,
\begin{equation}
\operatorname*{Im}({{\chi}_{\text{RI}}}) \approx \frac{2}{\pi} \frac{\epsilon_I}{\sqrt{\epsilon_R}}
\end{equation} 
as $\sin^{-1} (\frac{1}{\sqrt{\epsilon_R}}) \approx \frac{1}{\sqrt{\epsilon_R}}$ for $\epsilon_R \gg 1$. This reveals the contrast function becomes a straightforward ratio in this scenario.
For weak scattering, $\epsilon_R \approx 1, \delta \ll 1$, 
\begin{equation}
\operatorname*{Im}({{\chi}_{\text{RI}}}) \approx \epsilon_I
\end{equation} 
as $\sin^{-1} (\frac{1}{\sqrt{\epsilon_R}}) \approx \frac{\pi}{2}$ for $\epsilon_R\approx 1$. This is the conventional RA form as is expected.
 
We can substitute $\chi_{\text{RI}}$ (with averaged $\operatorname*{Im}({{\chi}_{\text{RI}}})$ in (\ref{Eq_avgIm1})) back to RI (\ref{Eq_RytovInt}) which provides the final form of the proposed method xPRA-LM, as
\begin{subequations}
	\label{Eq_RytovIntderived}
	\begin{align}
		P(\bm{r}) & [\text{dB}] = P_i(\bm{r})[\text{dB}]  \ + \nonumber \\ & C_0 \cdot \operatorname{Re}\bigg(\frac{k^2}{E_i(\bm{r})}  \int_{A} g(\bm{r}, \bm{r'})  \chi_{\text{RI}}(\bm{r'}) E_i(\bm{r'}) d\bm{r'}^2\bigg)
	\end{align}
	\begin{align}
		 	\operatorname*{Im}(\chi_{\text{RI}})(\bm{r}) & = \frac{2 \epsilon_I}{\pi} \sin^{-1} \biggl(\frac{1}{\sqrt{\epsilon_R}}\biggr)
	\end{align}
\end{subequations}

To summarize, we have used the high frequency approximation in low loss piece-wise homogeneous media to find an approximate expression for $\nabla \phi_s \cdot \nabla \phi_s$ which is  neglected from the contrast function of the conventional Rytov approximation. xPRA-LM is then based on including our expression for $\nabla \phi_s \cdot \nabla \phi_s$ into RI's contrast function. This provides a new expression for the reconstruction (\ref{Eq_RytovIntderived}) and is valid under a wide range of scattering conditions. To the best of our knowledge, this has not be performed before and as we shall show later it is remarkably accurate. 

\subsection{Background Subtraction and Imaging Temporal Change}
\label{sec_TBS}

To highlight a final feature of (\ref{Eq_RytovIntderived}) we also introduce the process of background subtraction and imaging temporal change. 

The indoor environment is unique \cite{Dubey2021} in that the stationary background (including ceiling, floor, walls, other clutter inside or outside DOI) will exhibit scattering. This is often significant and will cause distortion in the reconstruction. However these background objects will usually be spatially separate and distinct from the objects of interest. This observation implies that scattering or interaction between the objects of interest and the stationary background will usually not dominate the wave phenomena. Due to the linear form of (\ref{Eq_RytovIntderived}), subtraction of the background scattering can be included straightforwardly. 

Furthermore the objects of interest within the indoor environment are often those which are moving as in applications such as security and inventory control. Therefore subtracting out the stationary background component will leave behind the signal predominately relating only to those moving objects. Therefore background subtraction will significantly increase reconstruction accuracy of objects that have moved and remove distortion due to background scattering. We refer to this as imaging temporal change and it is performed by utilizing background subtraction.

Background subtraction also leads to the concept of temporal sparsity where changes in the indoor environment are usually minor compared to the background scatterers consisting of walls and furniture. We can therefore impose a strong temporal sparsity constraint on the change and we refer to this as temporal sparsity in the remainder of this paper. 

The use of temporal sparsity along with background subtraction are two powerful tools that can be utilized to enhance the performance of indoor imaging significantly. Due to the linear form of (\ref{Eq_RytovIntderived}), background subtraction can be included straightforwardly. This is in contrast to nonlinear techniques which cannot relate object change to measurement data change straightforwardly. 

Consider an initial time instant $t_0$ at which the contrast profile is $\chi_{\text{RI}}$ and the measured total power is $P^{t_0}$. Let there be some perturbation in this profile in duration $\Delta t$ so that the contrast profile at time instant $t_0+\Delta t$ is $\chi_{\text{RI}}^{t_0+\Delta t}$ and the measured total power is $P^{t_0+\Delta t}$. Writing (\ref{Eq_RytovIntderived}) for both time instances and subtracting them gives,
\begin{equation}
	\label{Eq_RIdBTBS}
	\begin{aligned}
		\Delta  & P(\bm{r}) [\text{dB}] = P^{t_0+\Delta t}(\bm{r}) - P^{t_0}(\bm{r})  \\ & = C_0 \cdot \operatorname{Re}\bigg(\frac{k^2}{E_i(\bm{r})}  \int_{A} g(\bm{r}, \bm{r'})  \Delta \chi_{\text{RI}}(\bm{r'}) E_i(\bm{r'}) d\bm{r'}^2\bigg)
	\end{aligned}
\end{equation}
where, 
\begin{equation}
	\label{deltaNu}
	\begin{aligned}
	\Delta \chi_{\text{RI}} &= \chi_{\text{RI}}^{t_0+\Delta t} - \chi_{\text{RI}}^{t_0}\\
	& = \Delta \operatorname*{Re}(\chi_{\text{RI}}) + j \Delta \operatorname*{Im}(\chi_{\text{RI}}).
	\end{aligned}
\end{equation}
This is the final form of xPRA-LM with background subtraction and it is extremely useful for imaging changes in an indoor environment, while removing distortions due to background scattering or clutter.

The form of xPRA-LM, (\ref{Eq_RIdBTBS}), is also very important for use in practical scenarios. When compared to (\ref{Eq_RytovIntderived}), we observe that we need to obtain estimates of the incident waveform. This is often difficult to obtain as an anechoic type environment would be needed to obtain these. However by leveraging background subtraction this complication is also removed. In essence, we acquire measurements when there are no objects in the DOI to obtain an estimate of the incident wavefield in the background clutter. Equating that to $t_0$ we then straightforwardly utilize (\ref{Eq_RIdBTBS}) by substituting $P^{t_0}=P^i$, $P^{t+\Delta t_0} = P, \chi_{\text{RI}}^{t_0}=0$ and $\chi_{\text{RI}}^{t+\Delta t_0} = \chi_{\text{RI}}$, where $P^i, P, \chi_{\text{RI}}$ are respectively the free-space incident power, total measured power in presence of scatterer and contrast of the scatterer.

\section{Simulation Results}
\label{Sec_Results}
In this section we provide simulation results to demonstrate the performance of the proposed xPRA-LM method in lossy media. In the section following, we provide experimental results. To make the simulated and experimental results compatible we make the geometry of DOI and sensor placement similar in both the simulation and experimental examples. 
For comparison, we have included the results for the real part of the contrast function $\operatorname*{Re}(\chi_{\text{RI}})$ in Appendix \ref{realpart}.  We have not provided comparisons to any other existing techniques because none of the other phaseless inverse scattering techniques can provide reconstructions for scatterers having large permittivity (up to $\epsilon_R=77$) and larger than $\lambda_0$ size. Most importantly, none of the techniques can be formulated for background subtraction which is crucial for practical indoor imaging. This has been explained in detail in section \ref{Sec_Intro}.

\begin{figure}[!ht]
	\centering
	\begin{subfigure}{0.24\textwidth}
		\includegraphics[width=\textwidth]{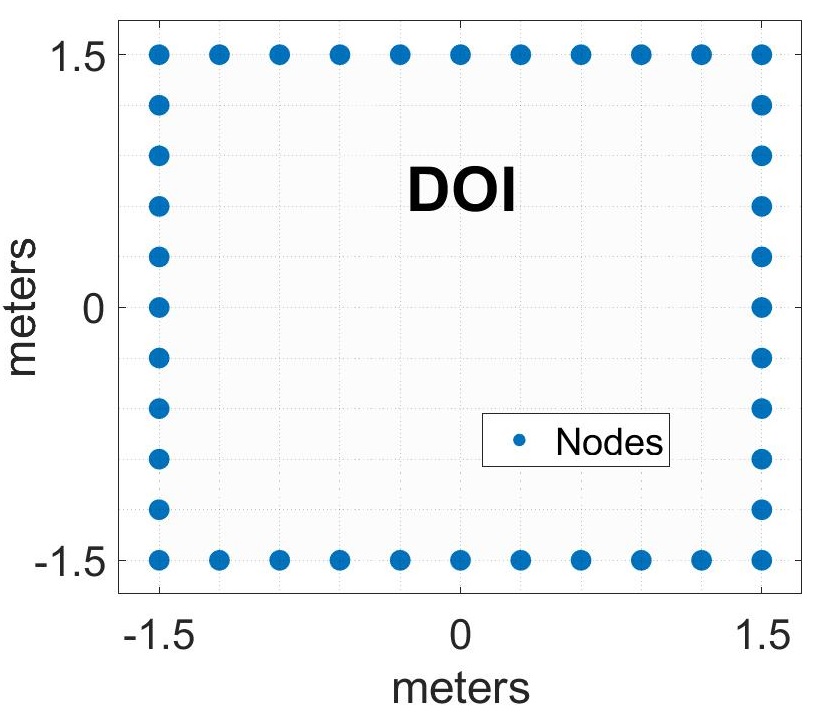}
		\subcaption{Simulation Setup}
	\end{subfigure}	
	\caption{Simulation setup showing $3 \times 3$ m$^2$ DOI with 40 transceiver nodes that can act as both sources and receivers of radiation.}
	\label{geometry} 
	\vspace{-0.5\baselineskip}		
\end{figure}

Fig. \ref{geometry} shows DOI setup for the simulations. The DOI size is $3 \times 3$ m$^2$ and transceiver nodes are placed at the boundary of the DOI. These can act as both sources and receivers of the wavefield to match the experimental configuration as detailed in the next section. The setup utilizes $M = 40$ transceiver nodes so that there are $L = \frac{M(M-1)}{2} = 780$ unique links. For reconstruction, the $3 \times 3$ m$^2$ DOI is divided into $2.5 \times 2.5$ cm$^2$ discrete grids (grid size in terms of wavelength = $\frac{\lambda_0}{5} \times \frac{\lambda_0}{5}$). Hence we need to estimate $N = 14400$ unknowns using $780$ measurements which gives a severely under-determined inverse problem and requires regularization. To perform regularization we use conventional 2D total variation (TV) regularization to obtain our reconstructions \cite{li2009user}.

To generate simulation data, we use the method of moments to obtain the forward simulation of scattering. Detailed equations for simulating the forward problem can be found in \cite{chen2018computational}. For the forward problem, the DOI is divided into discrete grids of size $\frac{\lambda}{10} \times \frac{\lambda}{10}$, where $\lambda = \lambda_0/\sqrt{\epsilon_R}$. Hence, the grid size in the forward problem is always smaller than the grid size used for the inverse problem and the method for calculating the forward problem is different from the inversion process to avoid inverse crimes \cite{chen2018computational}.

We consider the following configurations for numerical evaluation:
\begin{enumerate}
	\item Single large scatterer with $\delta \ll 1$, $\epsilon_R\approx 1$ \textbf{(Low-Loss, Weak scattering)}
	\item Single large scatterer with $\delta \ll 1$, $\epsilon_R > 1$ \textbf{(Low-Loss, Strong scattering)}
	\item Single large scatterer with $\delta \ll 1$, $\epsilon_R \gg 1$ \textbf{(Low-Loss, Extremely strong scattering)}	
	\item Multiple large scatterers with $\delta \ll 1$ and different $\epsilon_R$ values \textbf{(Low-Loss, Strong multiple scattering)}
	\item Scatterer with temporal change in profile ( $\delta \ll 1$ and $\epsilon_R$)	
\end{enumerate}

For all the results, we use scatterers which are larger than the incident wavelength (for example $\lambda_0 = 12.5 $ cm for 2.4 GHz WiFi). For reconstructing profiles, we only use phaseless data (magnitude only). To quantify the performance, we use the peak signal to noise ratio (PSNR) and Structural Similarity Index (SSIM) \cite{ssim2004} which are commonly used image quality assessment metric to compare the ground truth image to the reconstructed image. The higher the PSNR value, the better is the reconstruction. SSIM value ranges from 0 to 1 and the closer it is to 1, the closer is the reconstruction to the ground truth. Typically PSNR values greater than 20-30 dB are considered good while SSIM values greater than 0.85 are considered very good. The PSNR and SSIM values are provided in the figure captions of all reconstruction results.    

For better visualization, all the reconstructions are zoomed in on a $1 \times 1$ m$^2$ central area of the reconstruction to highlight the area around the scatterer rather than showing reconstruction of the whole $3 \times 3$ m$^2$ DOI. Also, in the remainder of this section, we denote the maximum physical length of the scatterer as $L_p$ and the corresponding maximum electrical length as $L_e$ which are expressed respectively in terms of free space wavelength $\lambda_0$ and wavelength inside the scatterer $\lambda = \lambda_0/\sqrt{\epsilon_R}$. Both $L_p$ and $L_e$ are measured along the longest linear length across the 2D cross section of the scatterer. For example, if the scatterer is square shaped, then $L_p$ and $L_e$ are measured along the diagonal.

\subsection{Numerical Examples and Analysis}
\label{Sec_NumExam}
\subsubsection{\textbf{Low-Loss, Weak scattering}}
\label{Sec_LLLRI}

\begin{figure}[h]
	\captionsetup[subfigure]{justification=centering}
	\centering
	\begin{subfigure}[t]{0.23\textwidth}
		\includegraphics[width=\textwidth]{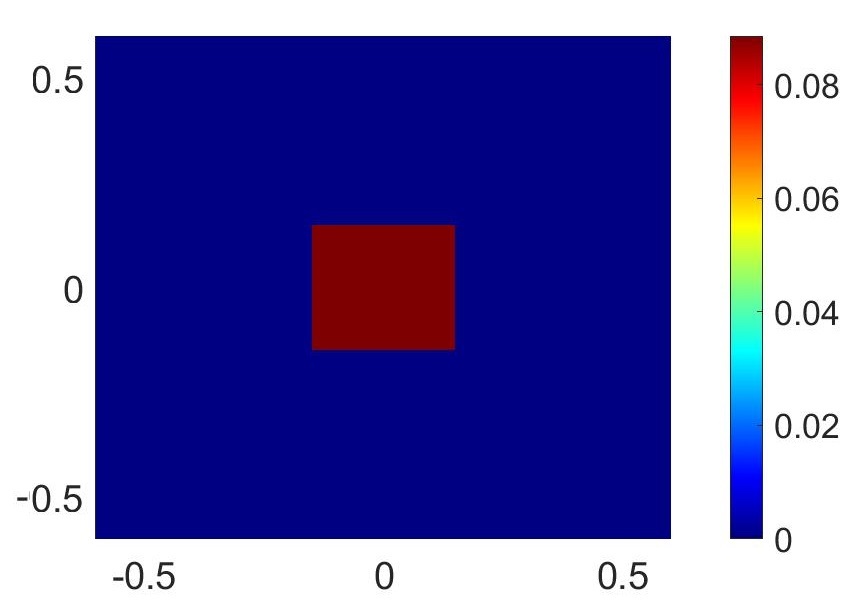}
		\subcaption{Ground Truth: $\operatorname*{Im}(\chi_{\text{RI}}) = 0.09$ where $\epsilon_R = 1.1, \epsilon_I = 0.11$.} 
	\end{subfigure}
	\begin{subfigure}[t]{0.23\textwidth}
		\includegraphics[width=\textwidth]{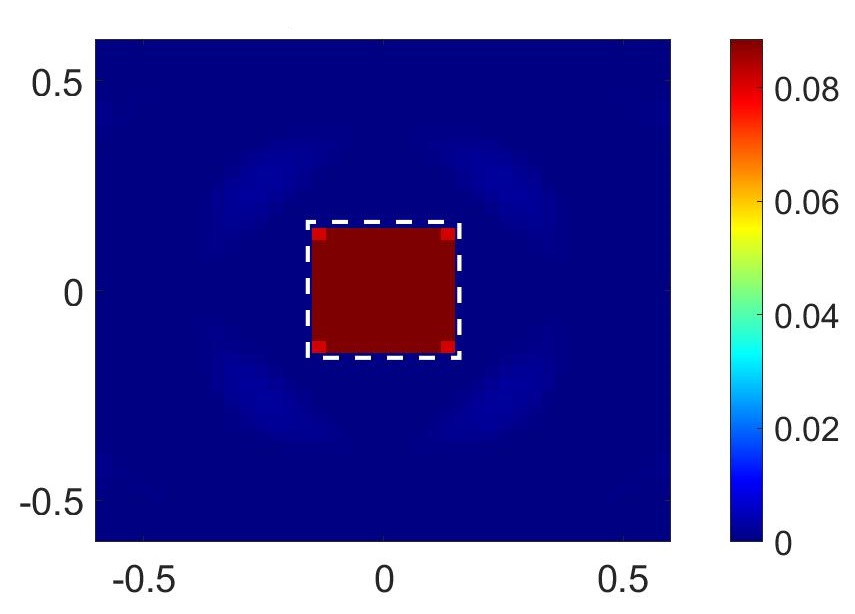}
		\subcaption{Reconstruction of $\operatorname*{Im}(\chi_{\text{RI}})$}
	\end{subfigure}	
	\caption{Reconstruction of imaginary component of contrast, $\operatorname*{Im}(\chi_{\text{RI}}) = \frac{2 \epsilon_I}{\pi} \sin^{-1}(1/\sqrt{\epsilon_R})$. The scatterer size is $30 \times 30$ cm$^2$ with $L_p = 3.4 \lambda_0$ and $L_e = 3.56 \lambda, \lambda = \lambda_0/\sqrt{1.1}$. (PSNR = 62 dB, SSIM = 0.998)}
	\label{LLLRI} 
	\vspace{-0.2\baselineskip}
\end{figure}

To verify the simulations we first provide a result for weak scattering. Fig. \ref{LLLRI} shows the reconstruction of a scatterer with $\epsilon_R = 1.1, \epsilon_I = 0.11, \delta = 0.1$. The scatterer size is $0.3 \times 0.3$ m$^2$ and hence, its physical length is $L_p = 3.4 \lambda_0$ and electrical length $L_e = 3.56 \lambda,$ where $\lambda = \lambda_0/\sqrt{\epsilon_R}$. From Fig. \ref{LLLRI}, it can be seen that the reconstructed imaginary part $\operatorname*{Im}(\chi_{\text{RI}}) = \frac{2 \epsilon_I}{\pi} \sin^{-1}(1/\sqrt{\epsilon_R})$ is an excellent match with the ground truth. Furthermore, the value of the reconstructed profile is also close to the expected results from conventional RA, i.e. $\frac{2 \epsilon_I}{\pi} \sin^{-1}(1/\sqrt{\epsilon_R}) \approx \epsilon_I$ (as $\epsilon_R\approx 1$, $ \sin^{-1}(1/\sqrt{\epsilon_R}) \approx \frac{\pi}{2}$). The reconstruction of the real part of contrast is shown in Appendix B in Fig. \ref{LLLRI_R}, which is also accurate as expected for weak scattering. This validates that our formulation of low-loss, high frequency Rytov approximation (xPRA-LM) reduces to the conventional results under the weak scattering conditions. Reconstruction quality is excellent with high PSNR and SSIM values as expected.   

\subsection{Low-Loss, Strong Scattering}
\label{Sec_LLHRI}
\begin{figure}[h]
	\captionsetup[subfigure]{justification=centering}
	\centering      
	\begin{subfigure}[t]{0.23\textwidth}
		\includegraphics[width=\textwidth]{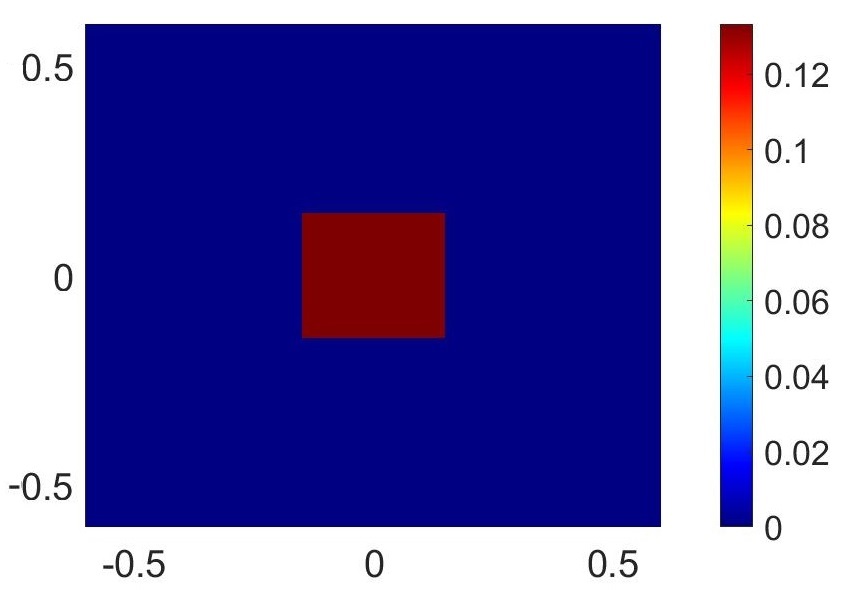}
		\subcaption{Ground Truth: $\operatorname*{Im}(\chi_{\text{RI}}) = 0.13$ where $\epsilon_R = 4, \epsilon_I = 0.4$.} 
	\end{subfigure}      
	\begin{subfigure}[t]{0.23\textwidth}
		\includegraphics[width=\textwidth]{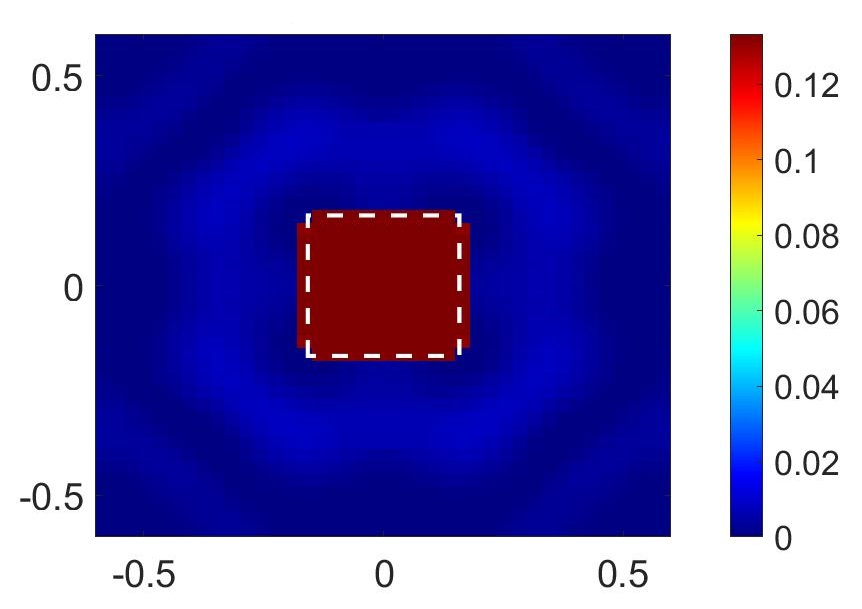}
		\subcaption{Reconstruction of $\operatorname*{Im}(\chi_{\text{RI}})$}
	\end{subfigure}	
	\caption{Reconstruction of imaginary component of contrast, $\operatorname*{Im}(\chi_{\text{RI}}) = \frac{2 \epsilon_I}{\pi} \sin^{-1}(1/\sqrt{\epsilon_R})$. The scatterer size is $30 \times 30$ cm$^2$ with $L_p = 3.4 \lambda_0$ and $L_e = 7 \lambda, \lambda = \lambda_0/\sqrt{4}$. (PSNR = 38 dB, SSIM = 0.981).}
	\label{LLHRIbooks} 
	\vspace{-0.2\baselineskip}			
\end{figure}

\begin{figure}[!h]
		\captionsetup[subfigure]{justification=centering}
			\centering
			\begin{subfigure}[t]{0.20\textwidth}
				\includegraphics[width=\textwidth]{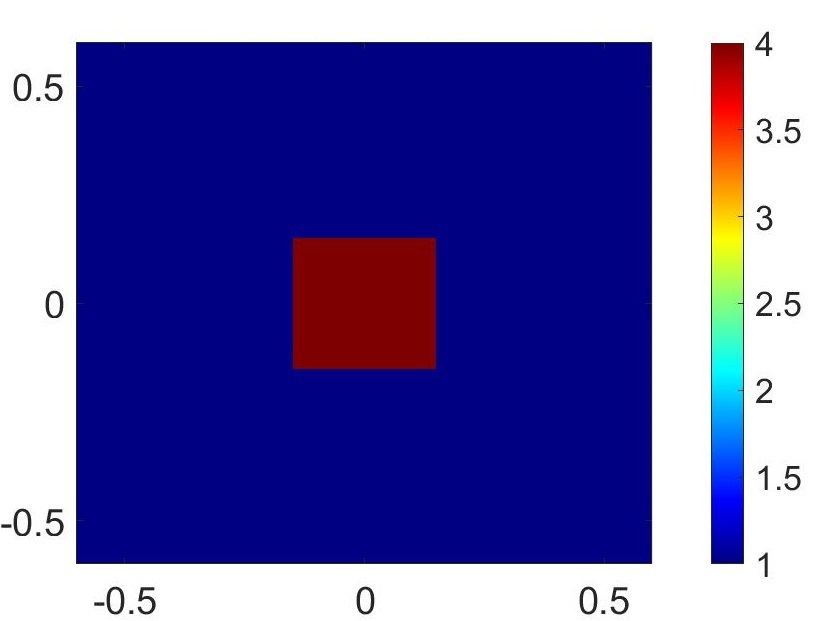}
				\subcaption{Ground Truth: $\epsilon_R = 4, \epsilon_I=0.4$.}
			\end{subfigure}       
			\begin{subfigure}[t]{0.21\textwidth}
				\includegraphics[width=\textwidth]{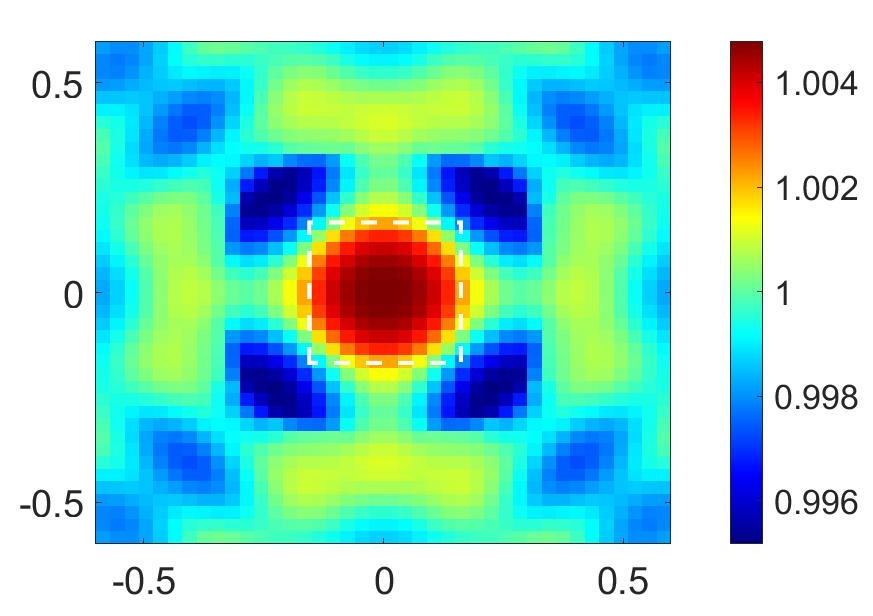}
				\subcaption{{Reconstruction of ${\epsilon_R}$}}
			\end{subfigure}       
			\caption{Real part reconstructions for profile shown in Fig. \ref{LLHRIbooks} (PSNR = 3.2 dB, SSIM=0.75).}			\label{LLHRIbooks_R} 
			\vspace{-0.5\baselineskip}			
\end{figure}

Next, we increase the permittivity further to $\epsilon_r = 4+j0.4$ (which is close to the permittivity of objects commonly found in indoor region such as bricks, concrete, wood and paper/books) \cite{4562803, ahmad2014partially, Productnote}. The scatterer has physical size $L_p = 3.4 \lambda_0$ and electrical size $L_e \approx 7 \lambda$ as shown in Fig. \ref{LLHRIbooks}. For conventional inverse scattering techniques, this permittivity and size is considered as very high. However, the reconstruction of the imaginary part of contrast $\operatorname*{Im}(\chi_{\text{RI}})$ in Fig. \ref{LLHRIbooks}(b) is surprisingly accurate even for this high permittivity and large size. PSNR and SSIM values are again very good. As expected, the reconstruction of  $\operatorname*{Re}(\chi_{\text{RI}})$ is very distorted (see Fig. \ref{LLHRIbooks_R}, Appendix \ref{realpart}) and the PSNR and SSIM values are both poor being less than 10 dB and 0.8 respectively.

\begin{figure}[h]
	\captionsetup[subfigure]{justification=centering}
	\centering     
	\begin{subfigure}[t]{0.23\textwidth}
		\includegraphics[width=\textwidth]{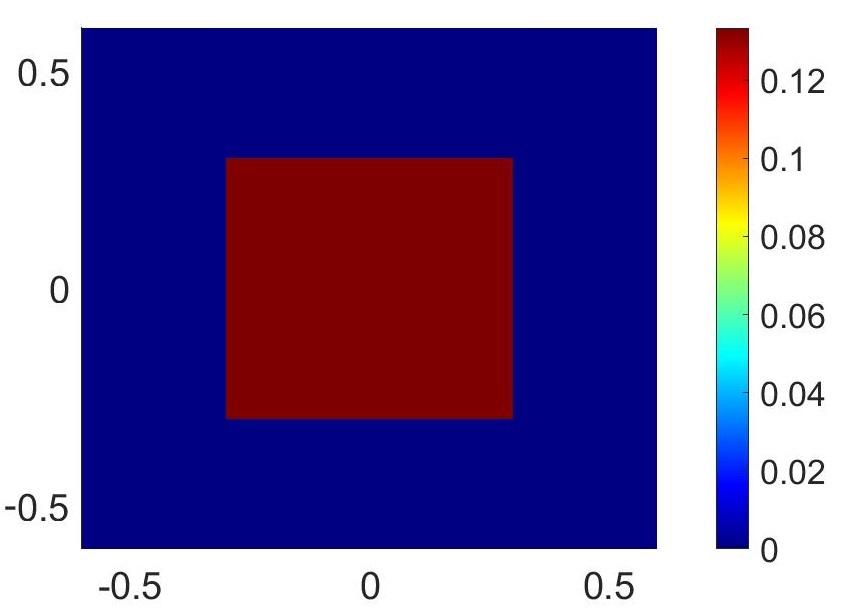}
		\subcaption{Ground Truth: $\operatorname*{Im}(\chi_{\text{RI}}) = 0.13$ where $\epsilon_R = 4, \epsilon_I = 0.4$.} 
	\end{subfigure}    
	\begin{subfigure}[t]{0.23\textwidth}
		\includegraphics[width=\textwidth]{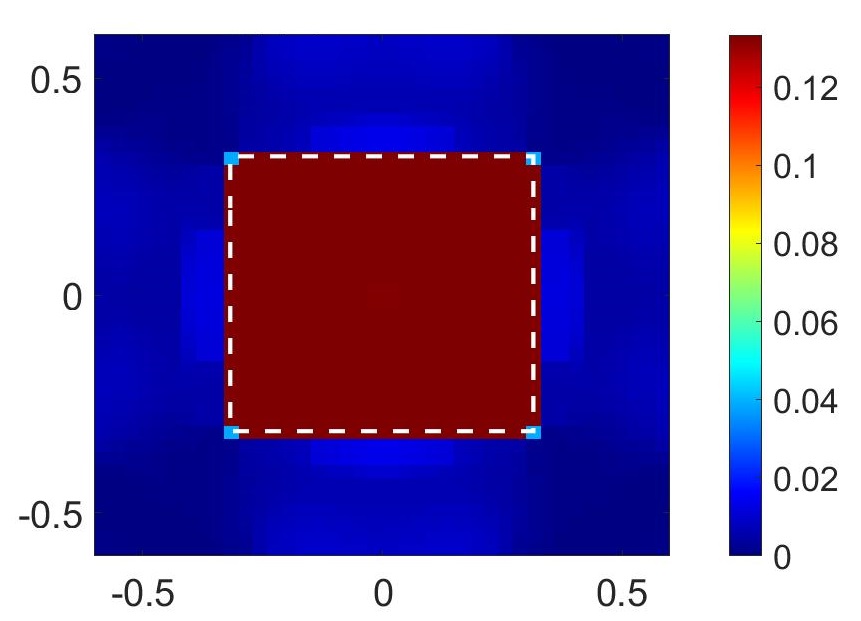}
		\subcaption{Reconstruction of $\operatorname*{Im}(\chi_{\text{RI}})$}
	\end{subfigure}	
	\caption{Reconstruction of imaginary component of contrast, $\operatorname*{Im}(\chi_{\text{RI}}) = \frac{2 \epsilon_I}{\pi} \sin^{-1}(1/\sqrt{\epsilon_R})$.  The scatterer size is $60 \times 60$ cm$^2$ with $L_p = 5 \lambda_0$ and $L_e = 10 \lambda, \lambda = \lambda_0/\sqrt{4}$. (PSNR = 42 dB, SSIM = 0.97).}
	\label{LLHRIbooks1} 
	\vspace{-0.2\baselineskip}			
\end{figure}
\begin{figure}[!h]
	\captionsetup[subfigure]{justification=centering}
	\centering      
	\begin{subfigure}[t]{0.23\textwidth}
		\includegraphics[width=\textwidth]{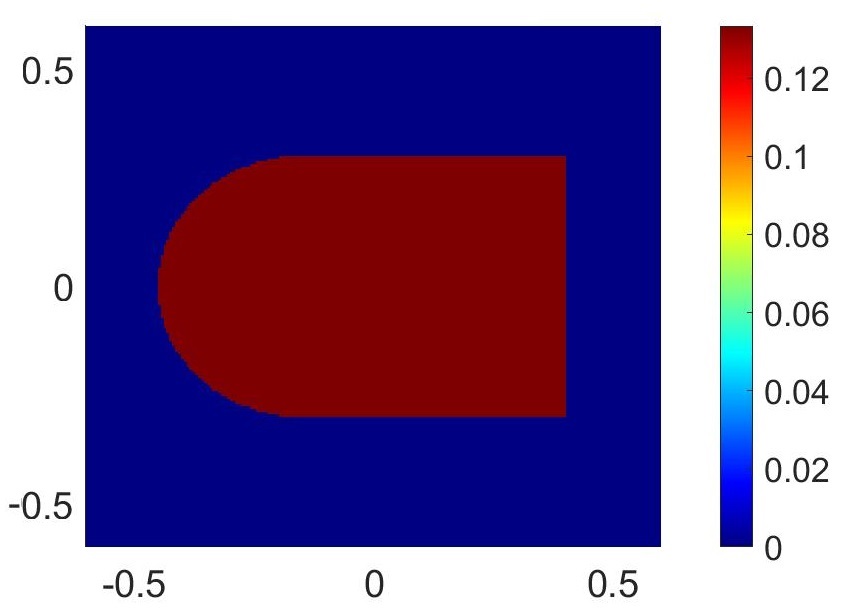}
		\subcaption{Ground Truth: $\operatorname*{Im}(\chi_{\text{RI}}) = 0.13$ where $\epsilon_R = 4, \epsilon_I = 0.4$.} 
	\end{subfigure}      
	\begin{subfigure}[t]{0.23\textwidth}
		\includegraphics[width=\textwidth]{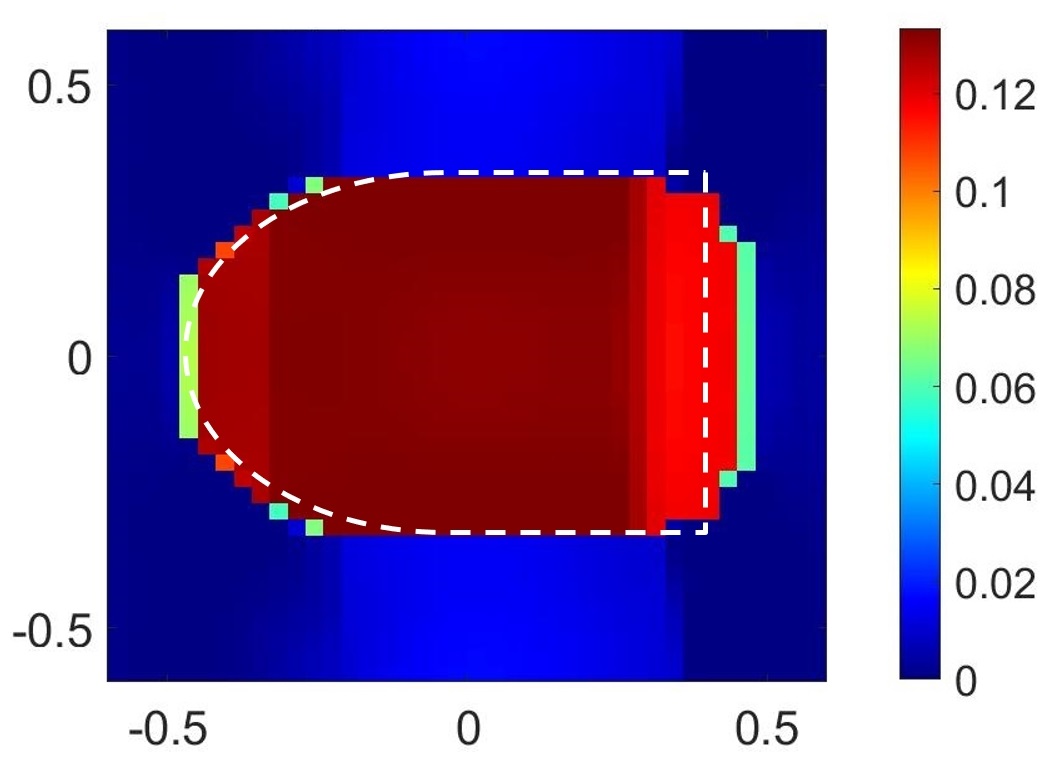}
		\subcaption{Reconstruction of $\operatorname*{Im}(\chi_{\text{RI}})$}
	\end{subfigure}	
	\caption{Reconstruction of imaginary component of contrast, $\operatorname*{Im}(\chi_{\text{RI}}) = \frac{2 \epsilon_I}{\pi} \sin^{-1}(1/\sqrt{\epsilon_R})$. (PSNR = 31 dB, SSIM = 0.9).}
	\label{LLHRIbooks2} 
	\vspace{-0.2\baselineskip}			
\end{figure}

Fig. \ref{LLHRIbooks1} and Fig. \ref{LLHRIbooks2} provide more examples with $\epsilon_r= 4+j0.4$ but with larger size compared to Fig. \ref{LLHRIbooks}. For the scatterer in Fig. \ref{LLHRIbooks1}, the physical and electrical sizes are $L_p = 6.8 \lambda_0$ and $L_e = 13.6 \lambda$ respectively and is twice the size used in Fig. \ref{LLHRIbooks}. Once, again, we see that the imaginary part of reconstruction is remarkably accurate. Similar conclusions can be seen for the results in Fig. \ref{LLHRIbooks2} for another large scatterer which is a different shape from that of the previous examples which consisted of square scatterers. PSNR and SSIM remain very good and are above 30 dB and 0.85 respectively.

\subsection{Low-Loss, Extremely Strong Scattering}
\label{Sec_LLHRIhuman}
\begin{figure}[h]
	\captionsetup[subfigure]{justification=centering}
	\centering      
	\begin{subfigure}[t]{0.23\textwidth}
		\includegraphics[width=\textwidth]{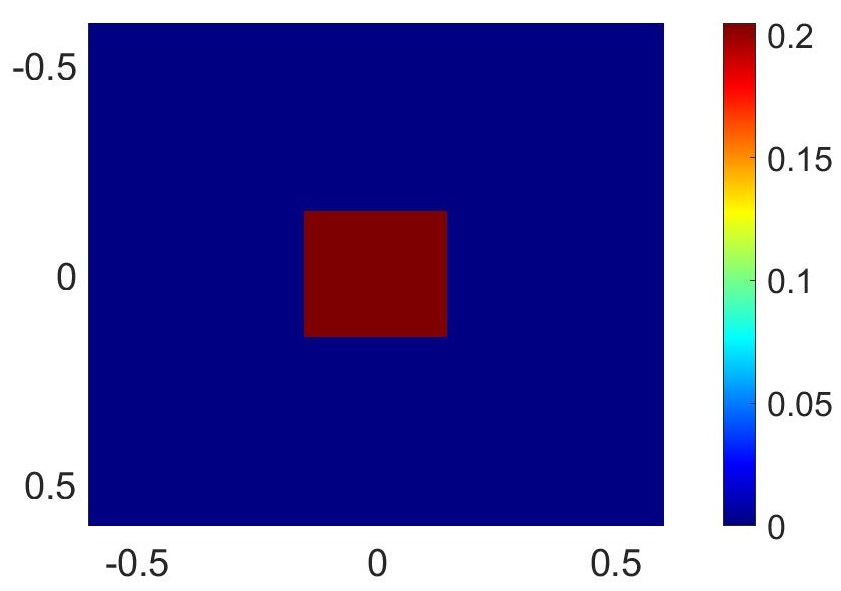}
		\subcaption{Ground Truth: $\operatorname*{Im}(\chi_{\text{RI}}) = 0.2$ where $\epsilon_R = 10, \epsilon_I = 1$.} 
	\end{subfigure}      
	\begin{subfigure}[t]{0.23\textwidth}
		\includegraphics[width=\textwidth]{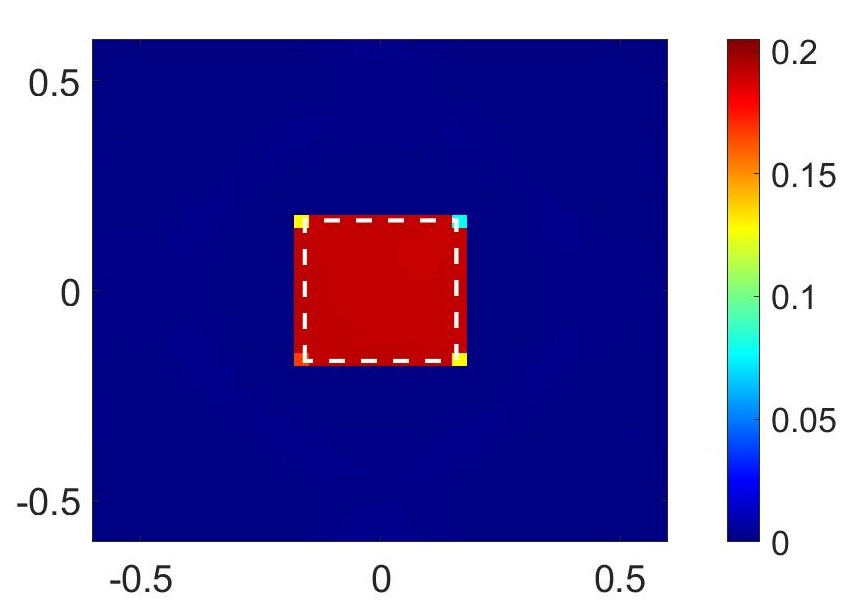}
		\subcaption{Reconstruction of $\operatorname*{Im}(\chi_{\text{RI}})$}
	\end{subfigure}	
	\caption{Reconstruction of imaginary component of contrast, $\operatorname*{Im}(\chi_{\text{RI}}) = \frac{2 \epsilon_I}{\pi} \sin^{-1}(1/\sqrt{\epsilon_R})$. The scatterer size is $30 \times 30$ cm$^2$ with $L_p = 3.4 \lambda_0$ and $L_e = 11 \lambda, \lambda = \lambda_0/\sqrt{10}$. (PSNR = 33 dB, SSIM = 0.97).}
	\label{LLHRIhuman10} 
	\vspace{-0.2\baselineskip}			
\end{figure}

\begin{figure}[h]
	\captionsetup[subfigure]{justification=centering}
	\centering      
	\begin{subfigure}[t]{0.23\textwidth}
		\includegraphics[width=\textwidth]{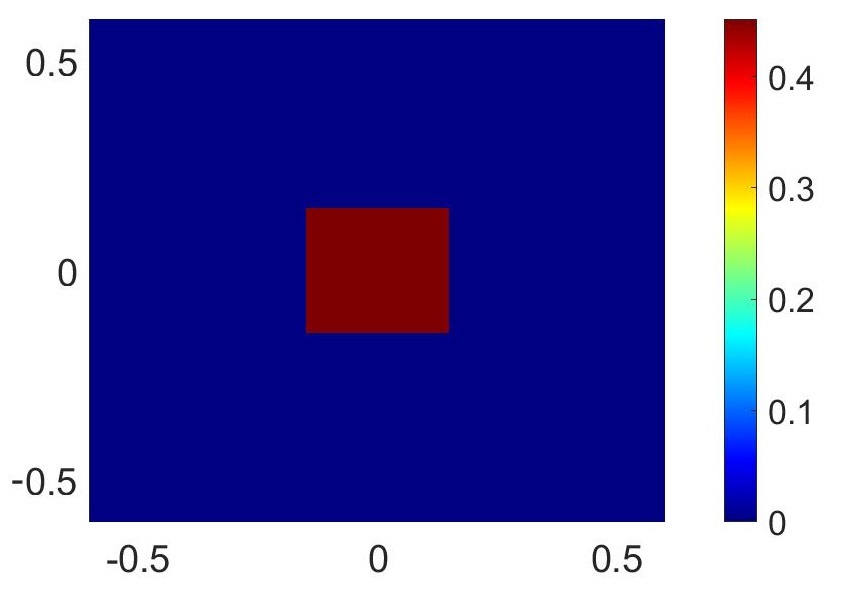}
		\subcaption{Ground Truth: $\operatorname*{Im}(\chi_{\text{RI}}) = 0.45$ where $\epsilon_R = 50, \epsilon_I = 5$.} 
	\end{subfigure}      
	\begin{subfigure}[t]{0.23\textwidth}
		\includegraphics[width=\textwidth]{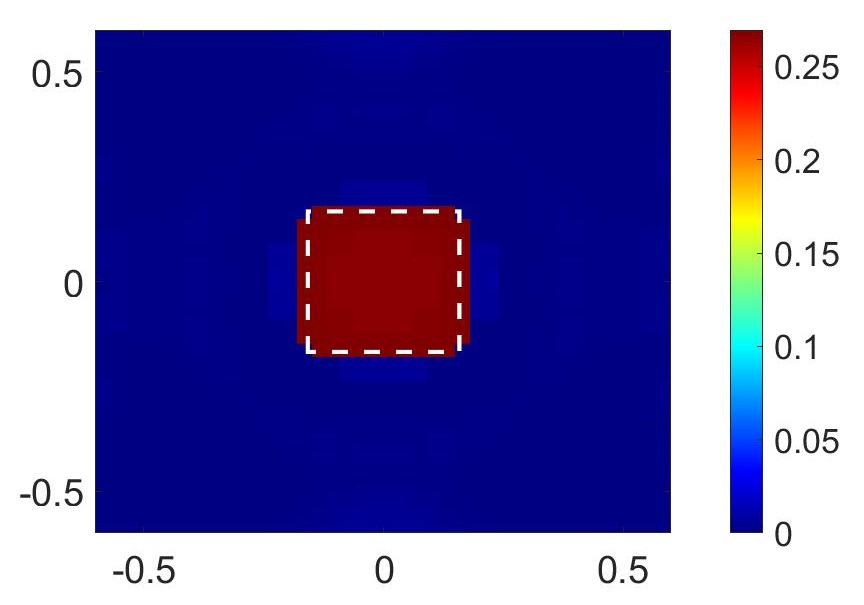}
		\subcaption{Reconstruction of $\operatorname*{Im}(\chi_{\text{RI}})$}
	\end{subfigure}	
	\caption{Reconstruction of imaginary component of contrast, $\operatorname*{Im}(\chi_{\text{RI}}) = \frac{2 \epsilon_I}{\pi} \sin^{-1}(1/\sqrt{\epsilon_R})$.  The scatterer size is $30 \times 30$ cm$^2$ with $L_p = 3.4 \lambda_0$ and $L_e = 24 \lambda, \lambda = \lambda_0/\sqrt{50}$. (PSNR = 25 dB, SSIM = 0.95).}
	\label{LLHRIhuman} 
	\vspace{-0.2\baselineskip}			
\end{figure}

\begin{figure}[!h]
	\captionsetup[subfigure]{justification=centering}
	\centering    
	\begin{subfigure}[t]{0.23\textwidth}
		\includegraphics[width=\textwidth]{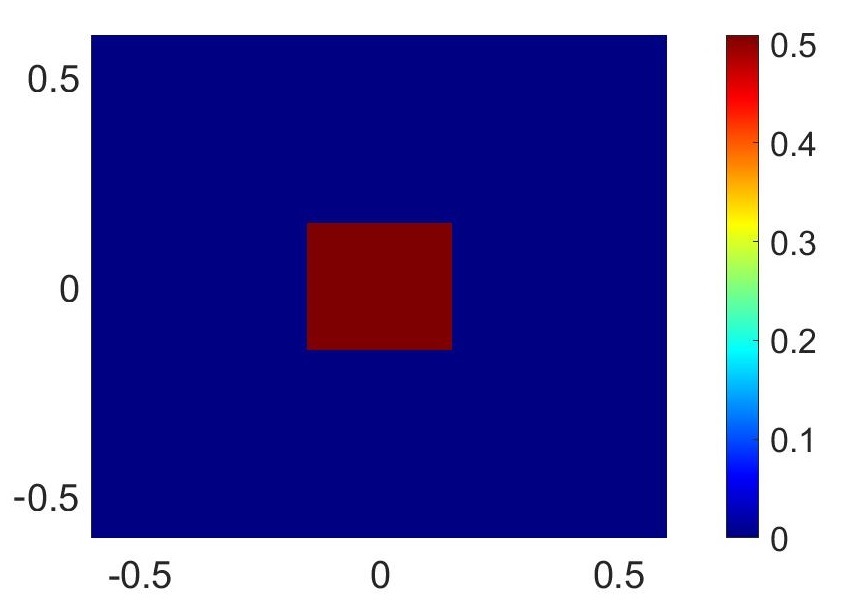}
		\subcaption{Ground Truth: $\operatorname*{Im}(\chi_{\text{RI}}) = 0.45$ where $\epsilon_R = 77, \epsilon_I = 7$.} 
	\end{subfigure}    
	\begin{subfigure}[t]{0.23\textwidth}
		\includegraphics[width=\textwidth]{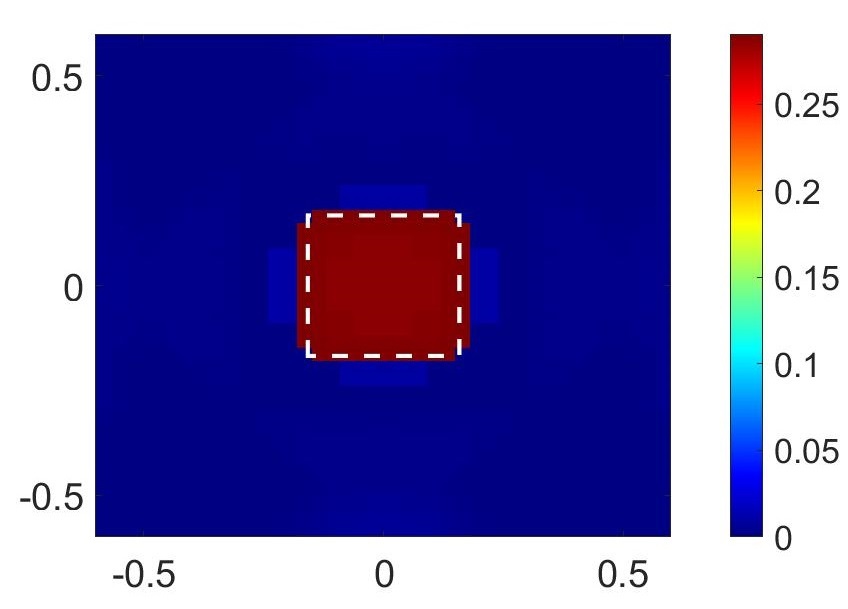}
		\subcaption{Reconstruction of $\operatorname*{Im}(\chi_{\text{RI}})$}
	\end{subfigure}	
	\caption{Reconstruction of imaginary component of contrast, $\operatorname*{Im}(\chi_{\text{RI}}) = \frac{2 \epsilon_I}{\pi} \sin^{-1}(1/\sqrt{\epsilon_R})$. The scatterer size is $30 \times 30$ cm$^2$ with $L_p = 3.4 \lambda_0$ and $L_e = 30 \lambda, \lambda = \lambda_0/\sqrt{77}$. (PSNR = 24 dB, SSIM = 0.942).}
	\label{LLHRIwater} 
	\vspace{-0.2\baselineskip}			
\end{figure}

Next we increase permittivity to extremely high values ($\epsilon_r = 10+j1, \ 50+j5$ and $\epsilon_r = 77+7j$) while keeping $\delta = 0.1$. The values $\epsilon_r = \ 50+j5$ and $\epsilon_r = 77+7j$ also approximate the complex permittivity of water and the human body at 2.4 GHz \cite{4562803, ahmad2014partially} and are therefore useful in practice too. Figure \ref{LLHRIhuman10}, Fig. \ref{LLHRIhuman} and Fig. \ref{LLHRIwater} provides the reconstructions where the electrical size $L_e$ for these examples are $11 \lambda$, $24 \lambda$ and $30 \lambda$ respectively. To the best of our knowledge, there has been no demonstration for such strong scattering for objects larger than a wavelength in the inverse scattering community.


It can be seen in Fig. \ref{LLHRIhuman10}, Fig. \ref{LLHRIhuman} and Fig. \ref{LLHRIwater} that the shape reconstruction for the imaginary part of the contrast is remarkably accurate. The reconstruction amplitude is also accurate in Fig. \ref{LLHRIhuman10}. For Fig. \ref{LLHRIhuman} and Fig. \ref{LLHRIwater}, the reconstruction amplitude however is slightly underestimated. This can be expected because of the extremely large values of $\epsilon_R, \epsilon_I$. In such a scenario, several assumptions used to derive xPRA-LM start to be violated. Also, the errors due to the approximations used to derive (\ref{Eq_avgIm1}) from (\ref{Eq_rytovfulldB6})) can also be amplified. Another reason relates to the averaging operation performed in (\ref{Eq_avgIm}) to make $\operatorname*{Im}(\chi_{\text{RI}})$ independent of $\theta_i$. The averaging is performed by assuming that the angle of incidence of the waves is uniform in the interval $\theta_i \in [-\pi/2, \pi/2]$ which is not completely satisfied and any deviations from this assumption can result in errors as the entities involved ($\epsilon_R, \epsilon_I$) are large in Fig. \ref{LLHRIhuman} and Fig. \ref{LLHRIwater}. On the other hand since $\cos\theta_i> 0 $ for any value of $\theta_i\in (-90^\circ, 90^\circ)$ and permittivity, the shape of the reconstruction of the imaginary part will not be distorted (as seen in Fig. \ref{LLHRIhuman}b and Fig. \ref{LLHRIwater}b) even if the averaging operation is not accurate. PSNR and SSIM remain good for both results and are above 20 dB and 0.85 respectively. In contrast the results shown in Appendix B have PSNR and SSIM values that are both poor being less than 10 dB and 0.8 respectively.

\subsection{Overall Summary}

\begin{figure}[!h]
	\centering
	\begin{subfigure}{0.35\textwidth}
		\includegraphics[width=\textwidth]{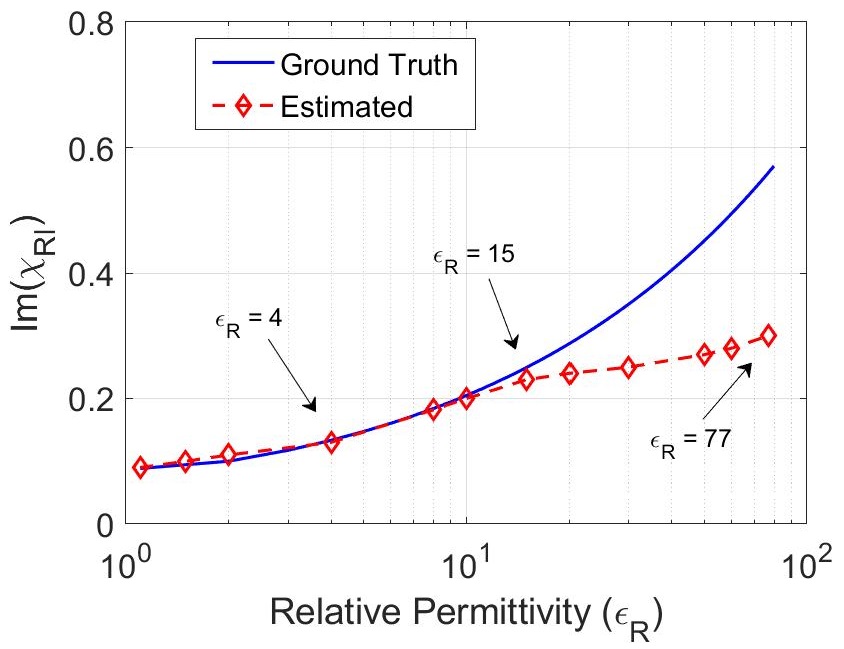}
	\end{subfigure}	      
	\caption{Plot in blue solid line represents ground truth amplitude of $\operatorname*{Im}(\chi_{\text{RI}}) = \frac{2 \epsilon_I}{\pi} \sin^{-1}(1/\sqrt{\epsilon_R})$ as a function of real part $\epsilon_R$ of relative permittivity. The value of loss tangent is fixed at $\delta=0.1$ so that the imaginary part of permittivity of scatterers becomes $0.1 \times \epsilon_R$. Plot in red dotted line represents reconstructed value of $\operatorname*{Im}(\chi_{\text{RI}})$ using xPRA-LM.}
	\label{ValueComparsion} 
	\vspace{-0.5\baselineskip}		
\end{figure}

Overall, results show that reconstruction of shape is accurate for any value of permittivity but estimation of contrast function amplitude is affected if permittivity is too large. To obtain an estimate of the reconstruction accuracy as a function of refractive index, we consider a circular object of diameter 5$\lambda_0$ with varying complex permittivities. Figure \ref{ValueComparsion} plots the ground truth verses the reconstructed amplitude value of the imaginary part of the contrast function ($\operatorname*{Im}(\chi_{\text{RI}})$) as a function of relative permittivity (where real part $\epsilon_R$ is varied in the range  $\epsilon_R \in [1.1, 80]$ while loss tangent is fixed at $\delta = 0.1$ so that $\epsilon_I$ is varied in the range $\epsilon_I \in [0.11, 8]$). It can be seen that the estimated value of $\operatorname*{Im}(\chi_{\text{RI}})$ given by xPRA-LM is accurate for a wide range of permittivity (it is accurate within 5\% error up to relative permittivity $\epsilon_r = 15 + j1.5$). 

To summarize, the results shown in Fig. \ref{LLHRIbooks} to Fig. \ref{LLHRIwater} show that the imaginary part of reconstruction of the contrast function $\operatorname*{Im}(\chi_{\text{RI}})$ provided by xPRA-LM achieve excellent shape estimation of the scatterers even for extremely high permittivity and large object sizes. The technique can therefore be very useful for indoor imaging applications such as imaging or tracking movement of objects or people. Apart from accurate shape reconstruction, xPRA-LM also provides accurate estimation of the value of $\operatorname*{Im}(\chi_{\text{RI}})$ which is accurate up to $\epsilon_r = 15 + j1.5$. Future work can include using iterative frameworks to first estimate the shape and then estimate $\cos\theta_i$ to compensate for the underestimated contrast function amplitudes by removing the need for uniform averaging with respect to $\theta_i$. Next we show that xPRA-LM can image multiple scatterers with different permittivity and can distinguish between these objects. 


\subsection{Multiple Scatterers}
\label{Sec_Multiple}
\begin{figure}[!h]
\captionsetup[subfigure]{justification=centering}
\centering  
	\begin{subfigure}[t]{0.23\textwidth}
	\includegraphics[width=\textwidth]{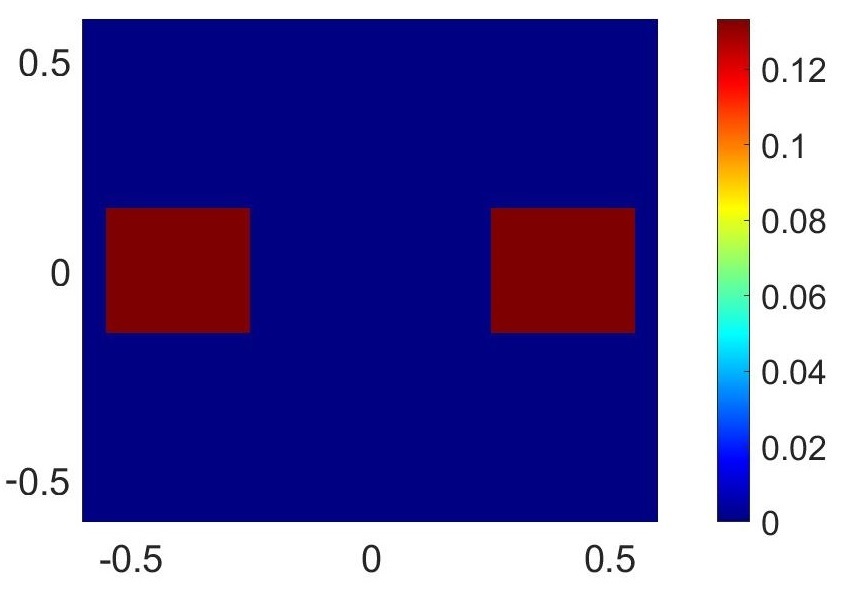}
		\subcaption{Ground Truth: $\operatorname*{Im}(\chi_{\text{RI}}) = 0.13$ where $\epsilon_R = 4, \epsilon_I = 0.4$.} 
	\end{subfigure}  
	\begin{subfigure}[t]{0.237\textwidth}
	\includegraphics[width=\textwidth]{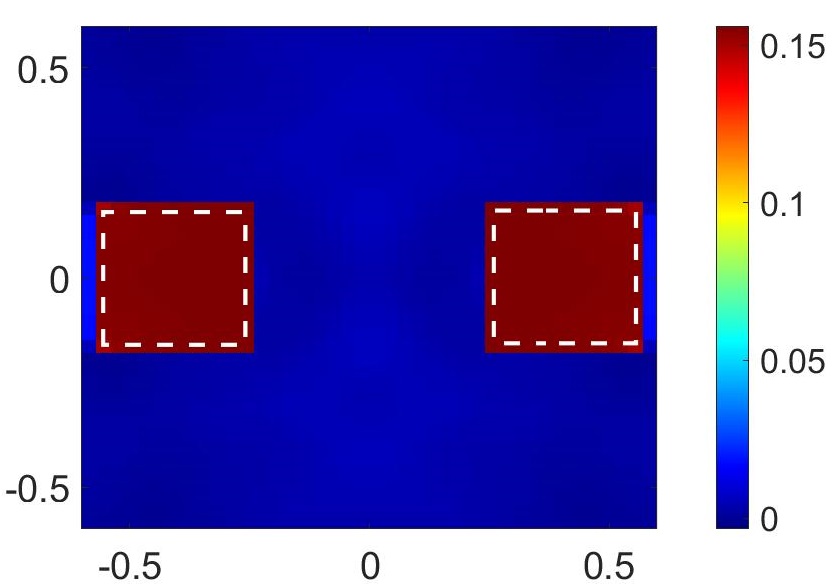}
		\subcaption{Reconstruction of $\operatorname*{Im}(\chi_{\text{RI}})$}
	\end{subfigure}	
	\caption{Reconstruction of imaginary component of contrast, $\operatorname*{Im}(\chi_{\text{RI}}) = \frac{2 \epsilon_I}{\pi} \sin^{-1}(1/\sqrt{\epsilon_R})$. The size of both scatterers is $30 \times 30$ cm$^2$ with $L_p = 3.4 \lambda_0$ and $L_e = 24 \lambda, \lambda = \lambda_0/\sqrt{50}$. (PSNR = 32 dB, SSIM = 0.98).}
\label{LLHRI_multiple1} 
\vspace{-0.2\baselineskip}			
\end{figure}

Fig. \ref{LLHRI_multiple1} provides the reconstruction of two scatterers (both with $\epsilon_r = 4 + j 0.4$) inside DOI. The reconstructions shows similar performance as for the configuration of a single scatterer and xPRA-LM provides excellent reconstruction of the scatterers. This shows that xPRA-LM can also handle multiple scattering between different scatterers. 

\begin{figure}[!h]
\captionsetup[subfigure]{justification=centering}
\centering      
	\begin{subfigure}[t]{0.23\textwidth}
	\includegraphics[width=\textwidth]{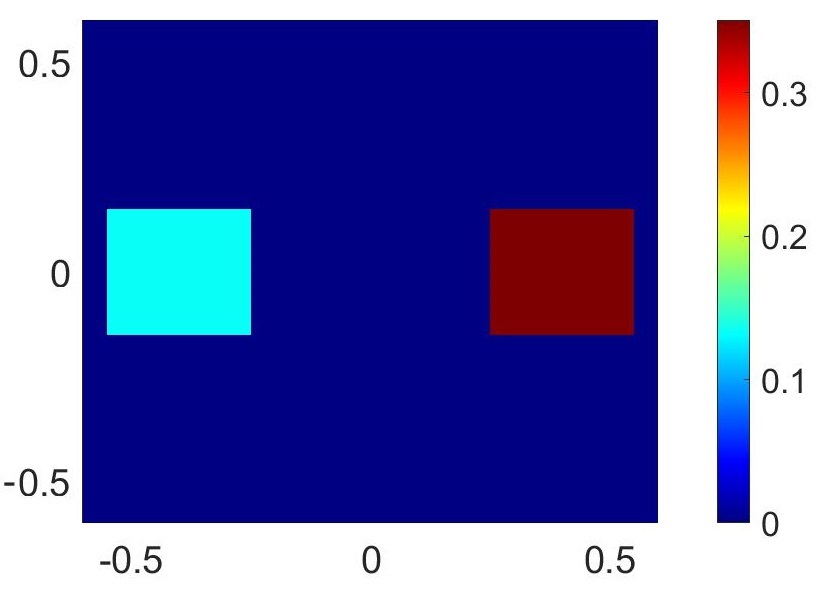}
		\subcaption{Ground Truth: $\operatorname*{Im}(\chi_{\text{RI}})$} 
	\end{subfigure}      
	\begin{subfigure}[t]{0.23\textwidth}
	\includegraphics[width=\textwidth]{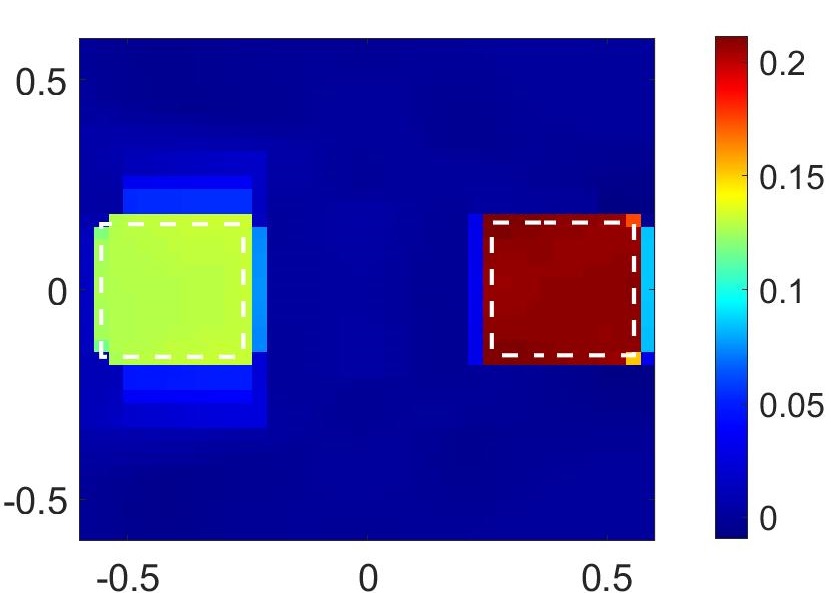}
		\subcaption{Reconstruction of $\operatorname*{Im}(\chi_{\text{RI}})$}
	\end{subfigure}	
	\caption{Reconstruction of imaginary component of contrast, $\operatorname*{Im}(\chi_{\text{RI}}) = \frac{2 \epsilon_I}{\pi} \sin^{-1}(1/\sqrt{\epsilon_R})$. The permittivity of scatterer on left is $\epsilon_R = 4, \epsilon_I = 0.4$ which gives $\operatorname*{Im}(\chi_{\text{RI}}) = 0.13$. The permittivity of scatterer on right is $\epsilon_R = 30, \epsilon_I = 3$  which gives $\operatorname*{Im}(\chi_{\text{RI}}) = 0.35$. The size of both scatterers is $30 \times 30$ cm$^2$. (PSNR = 22 dB, SSIM =  0.92).}
\label{LLHRI_multiple2} 
\vspace{-0.5\baselineskip}			
\end{figure}

\begin{figure}[!h]
\captionsetup[subfigure]{justification=centering}
\centering      
	\begin{subfigure}[t]{0.23\textwidth}
	\includegraphics[width=\textwidth]{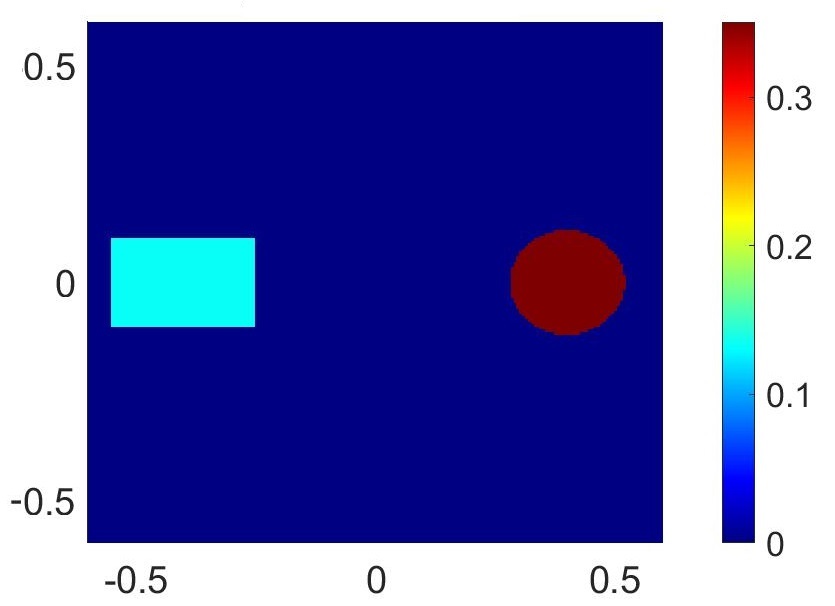}
		\subcaption{Ground Truth: $\operatorname*{Im}(\chi_{\text{RI}})$} 
	\end{subfigure}  
	\begin{subfigure}[t]{0.23\textwidth}
	\includegraphics[width=\textwidth]{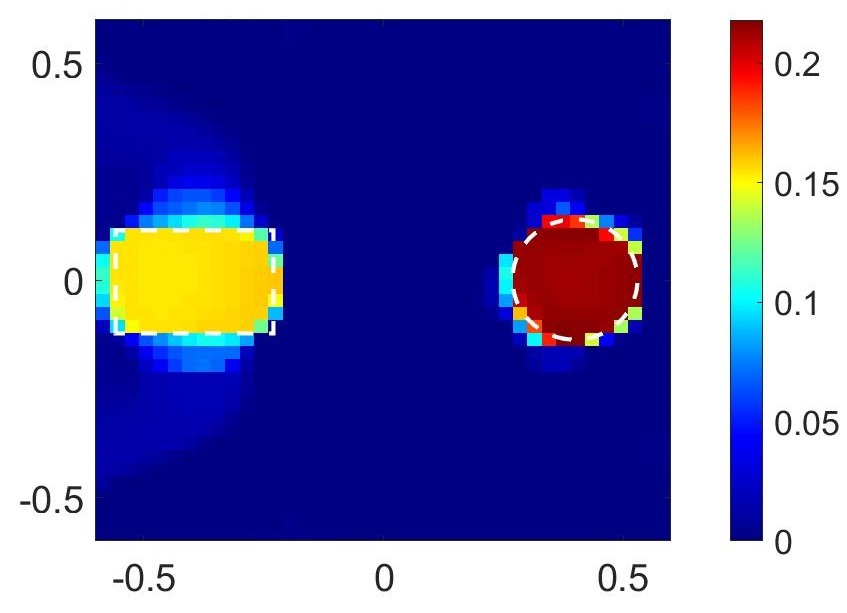}
		\subcaption{Reconstruction of $\operatorname*{Im}(\chi_{\text{RI}})$}
	\end{subfigure}	
	\caption{Reconstruction of imaginary component of contrast, $\operatorname*{Im}(\chi_{\text{RI}}) = \frac{2 \epsilon_I}{\pi} \sin^{-1}(1/\sqrt{\epsilon_R})$. The permittivity of scatterer on left is $\epsilon_R = 4, \epsilon_I = 0.4$ which gives $\operatorname*{Im}(\chi_{\text{RI}}) = 0.13$. The permittivity of scatterer on right is $\epsilon_R = 30, \epsilon_I = 3$  which gives $\operatorname*{Im}(\chi_{\text{RI}}) = 0.35$. (PSNR = 20 dB, SSIM =  0.9).}
\label{LLHRI_multiple3} 
\vspace{-0.5\baselineskip}			
\end{figure}

Fig. \ref{LLHRI_multiple2} shows reconstruction for two scatterers of different complex permittivity ($\epsilon_r = 4+ j 0.4$ for object on left and $\epsilon_r = 30 + j 3$ for object on right). This is an important test of the proposed xPRA-LM model because we desire shape reconstruction as well as being able to distinguish between different objects (reconstruction should be able to differentiate objects with different permittivity values). Fig. \ref{LLHRI_multiple2}(b) shows that xPRA-LM provides excellent shape reconstruction. The reconstructed image is also able to provide a clear distinction between the two objects. PSNR and SSIM remain good. 

Finally, Fig. \ref{LLHRI_multiple3} provides results where the two scatterers have different complex permittivity as well as different shapes. The conclusions are the same as Fig. \ref{LLHRI_multiple2} and reconstruction is accurate and provide clear distinction between the materials of the two scatterers. PSNR and SSIM remain good and in contrast the results shown in Appendix B have PSNR and SSIM values that are both poor being less than 10 dB and 0.8 respectively.

\subsection{Temporal Background Subtraction}
\label{TBS}
\begin{figure}[h]
	\captionsetup[subfigure]{justification=centering}
	\centering
	\begin{subfigure}[t]{0.13\textwidth}
	\includegraphics[width=\textwidth]{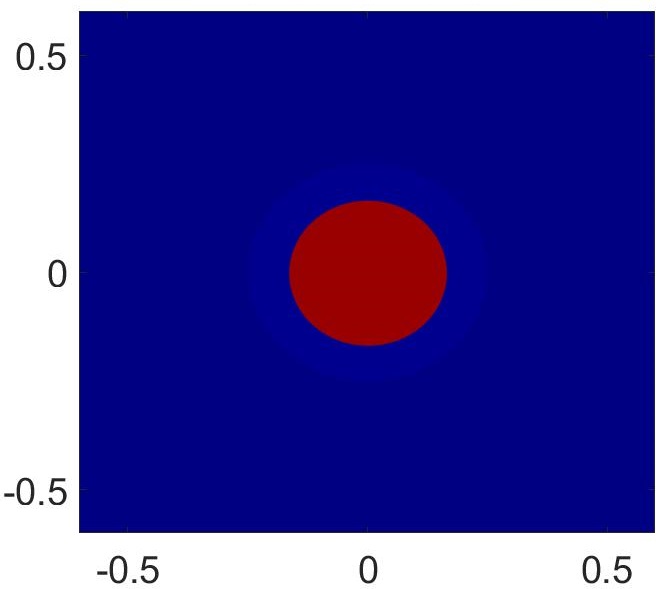}
	\subcaption{Profile in time $t=t_0$}
	\end{subfigure}       
	\begin{subfigure}[t]{0.13\textwidth}
	\includegraphics[width=\textwidth]{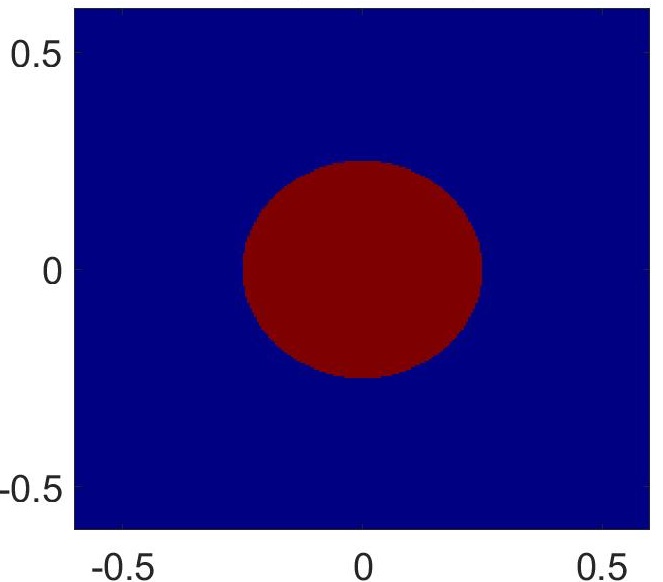}
	\subcaption{Profile in time $t=t_0+\Delta t$}
	\end{subfigure}
	\begin{subfigure}[t]{0.13\textwidth}
	\includegraphics[width=\textwidth]{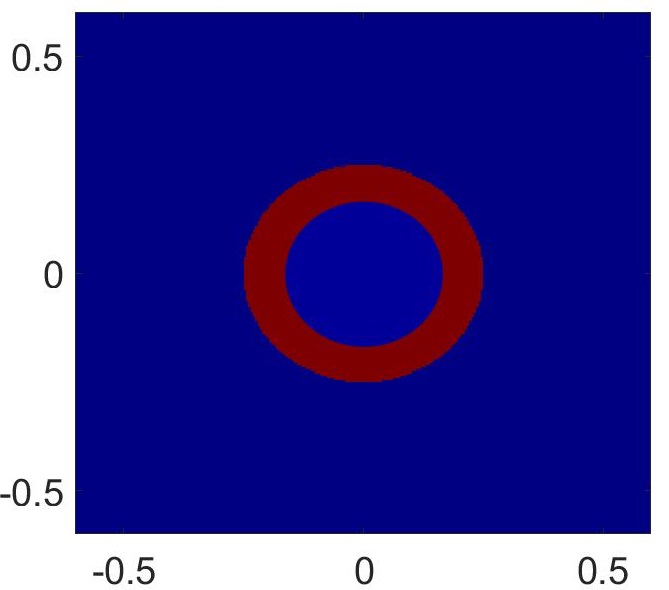}
	\subcaption{Change in profile $\Delta t$}
	\end{subfigure}       
	\caption{Temporal Background Subtraction}
	\label{LLHRI_TBSset} 
	\vspace{-0.5\baselineskip}	
\end{figure}

Next we examine the accuracy of xPRA-LM in imaging changes in the DOI profile and its efficiency in performing background subtraction. This is a crucial feature required for handling experimental data where measurements contain unwanted multipath distortions due to the stationary background such as ceiling, floor and walls. We selected the profile example shown in Fig. \ref{LLHRI_TBSset} as a demonstration. Fig. \ref{LLHRI_TBSset}(a) shows a circular profile at time instant $t_0$ which is changed to the  larger circular profile in Fig. \ref{LLHRI_TBSset}(b) after some time duration $\Delta t$. Therefore, the change in profile in time $\Delta t$ should appear as a ring as shown in Fig. \ref{LLHRI_TBSset}(c). 
	
\begin{figure}[!h]
\captionsetup[subfigure]{justification=centering}
	\centering      
	\begin{subfigure}[t]{0.23\textwidth}
		\includegraphics[width=\textwidth]{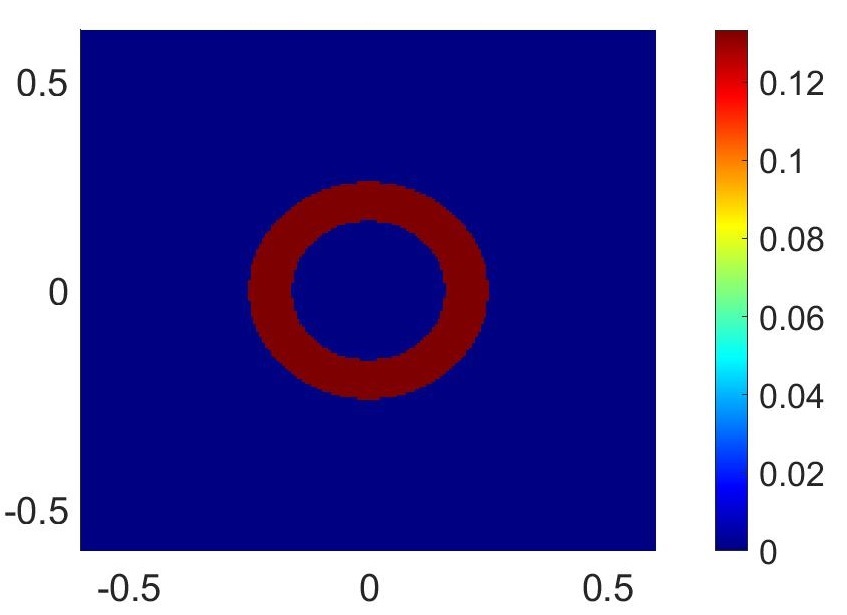}
		\subcaption{Ground Truth: $\operatorname*{Im}(\Delta \chi_{\text{RI}}) = 0.13$}
	\end{subfigure}	\begin{subfigure}[t]{0.23\textwidth}
		\includegraphics[width=\textwidth]{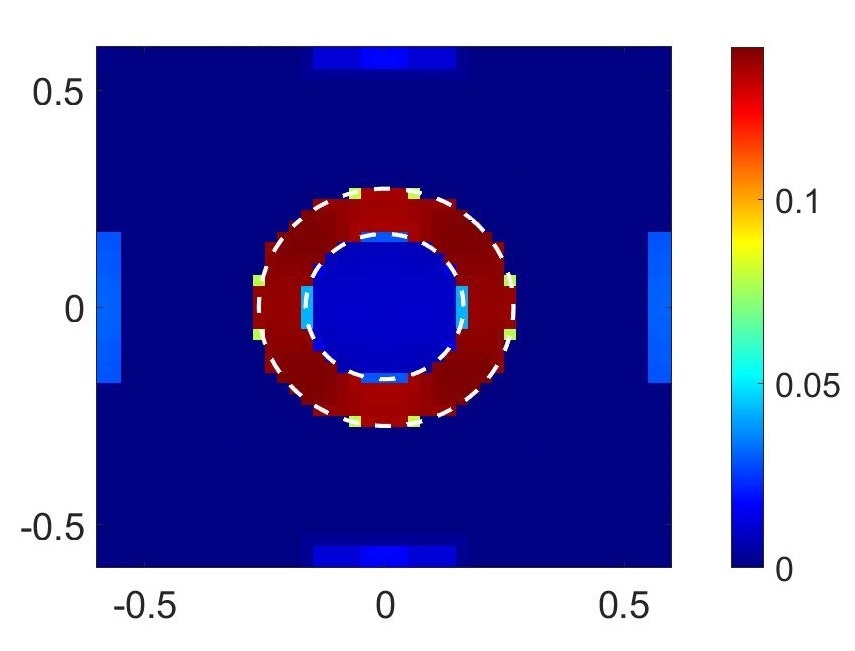}
		\subcaption{Reconstruction of $\operatorname*{Im}(\Delta \chi_{\text{RI}})$}
	\end{subfigure}	
	\caption{Demonstrating Background Subtraction results using xPRA-LM (PSNR = 33 dB, SSIM = 0.913)}
	\label{LLHRI_TBS} 
	\vspace{-0.5\baselineskip}
\end{figure}

We assign permittivity $\epsilon_r = 4 + j 0.4$ to the profile (change in profile) and the ground truth is shown in Fig. \ref{LLHRI_TBS}(a). Let the measured power at $t_0$ be $P^{t_0}$ and at $t_0+\Delta t$ be $p^{t_0+\Delta t}$. We substitute these in xPRA-LM (\ref{Eq_RIdBTBS}) to perform background subtraction and reconstruct the change in profile which should appear as a ring with $\operatorname*{Im}(\Delta \chi_{\text{RI}}) = 0.13$ (estimated using (\ref{deltaNu})). Fig. \ref{LLHRI_TBS}(b) shows that xPRA-LM is effective in reconstructing this change in profile.

\section{Experimental Results}
\label{Sec_exp}
\begin{figure}[!h]
	\centering
	\begin{subfigure}{0.43\textwidth}
		\includegraphics[width=\textwidth]{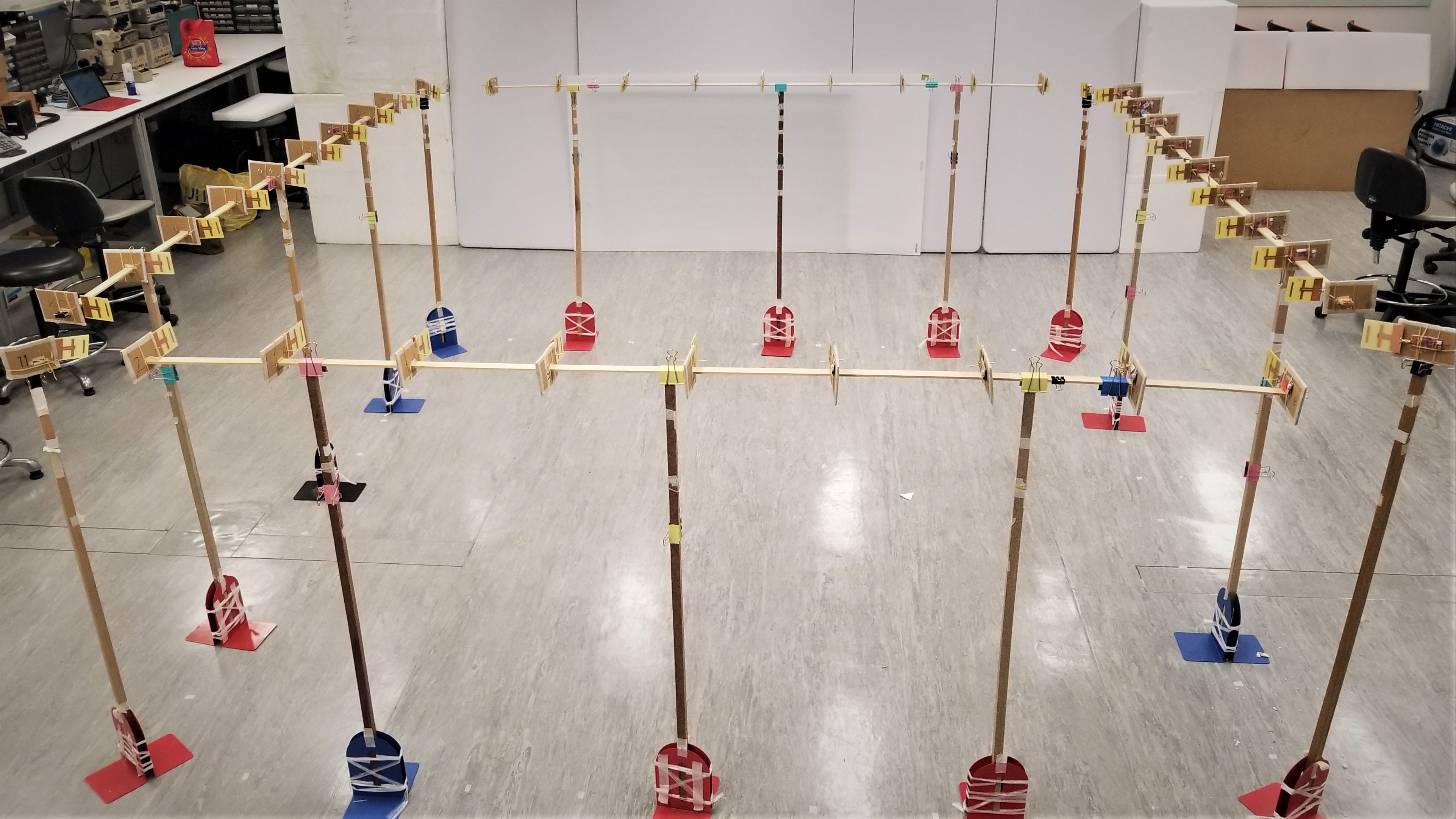}
	\end{subfigure}       
	\caption{Experimental setup showing $3 \times 3$ m$^2$ DOI with 40 WiFi 2.4 GHz transceiver nodes.}
	\label{geometryexp} 
	\vspace{-0.3\baselineskip}		
\end{figure}

Fig. \ref{geometryexp} shows our experimental setup. The DOI is inside room 3125A in the Hong Kong University of Science and Technology (HKUST). WiFi transceiver nodes can be seen at the boundary of the $3\times 3$ m$^2$ DOI. Each node consists of a SparkFun ESP32 Thing board \cite{ESP3200} consisting of an integrated 802.11 bgn WiFi transceiver operating at 2.4 GHz. The inbuilt omnidirectional antenna of the SparkFun ESP32 boards are replaced by a Yagi antenna of 6.6 dBi. The ESP32 boards (with antenna) are located at a height of $d_h=1.2$ m from the floor using a wooden stand (see Fig. \ref{geometryexp}). The Yagi antennas on the boards are oriented such that the xz-plane in Fig. \ref{geometryexp} lies in the 2D DOI plane and yz-plane (center dipole element of the antenna) is normal to 2D DOI plane and hence matches our TM (vertical polarization) simulations and formulations. Every transceiver can be assumed to alternate between access point (AP) and station (STA) mode so that the WiFi beacon signal can be utilized to obtain the RSSI for each link. This experimental configuration utilizes $M=40$ WiFi transceiver nodes so that there are in total $L=780$ unique measurement links ($L=M(M-1)/2$). 
 
\begin{figure*}[h]
	\captionsetup[subfigure]{justification=centering}
	\centering
	\begin{subfigure}[t]{0.34\textwidth}
		\includegraphics[width=\textwidth]{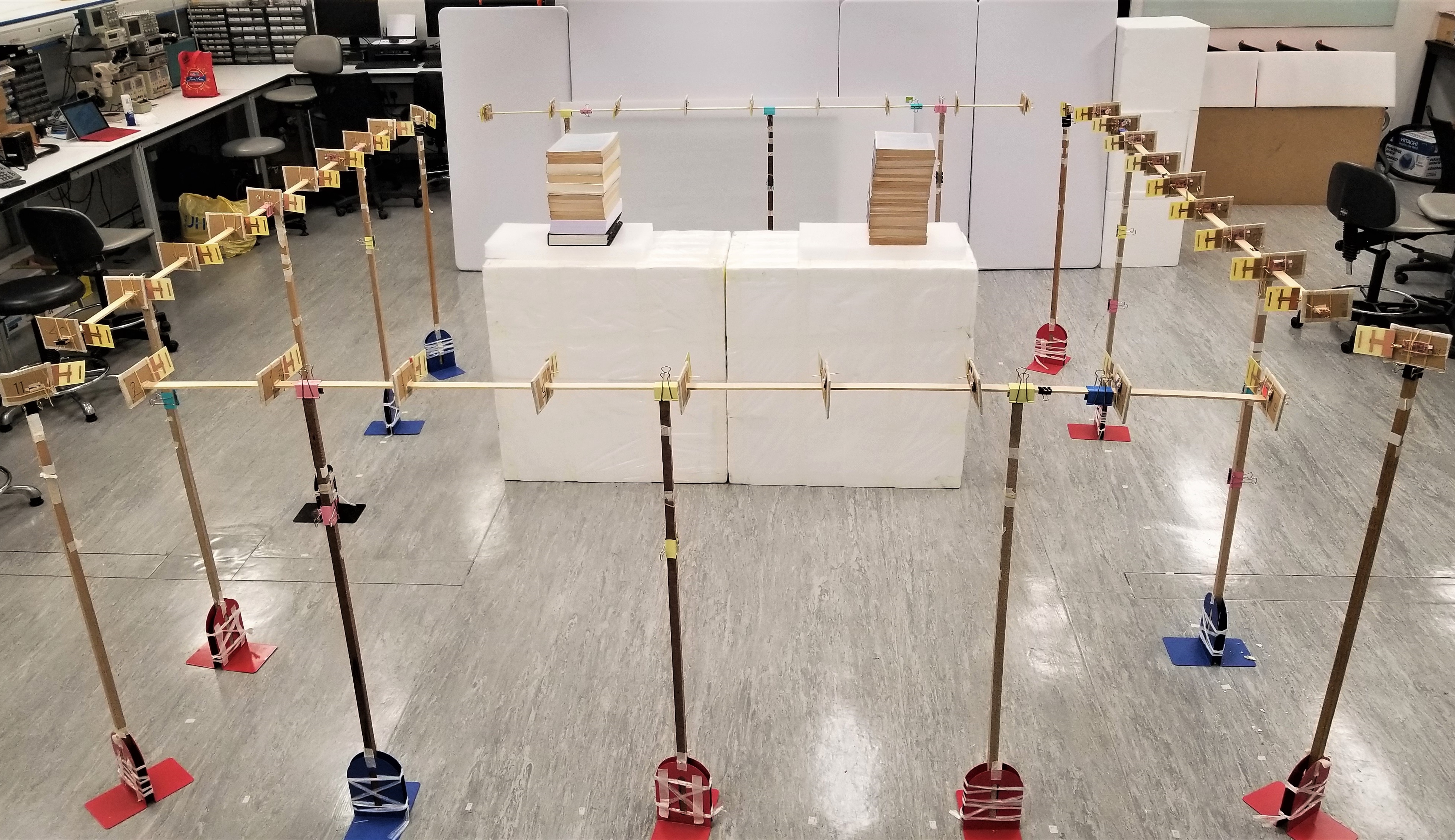}
		\subcaption{Experimental Setup}
	\end{subfigure} 	   
	\begin{subfigure}[t]{0.2\textwidth}
		\includegraphics[width=\textwidth]{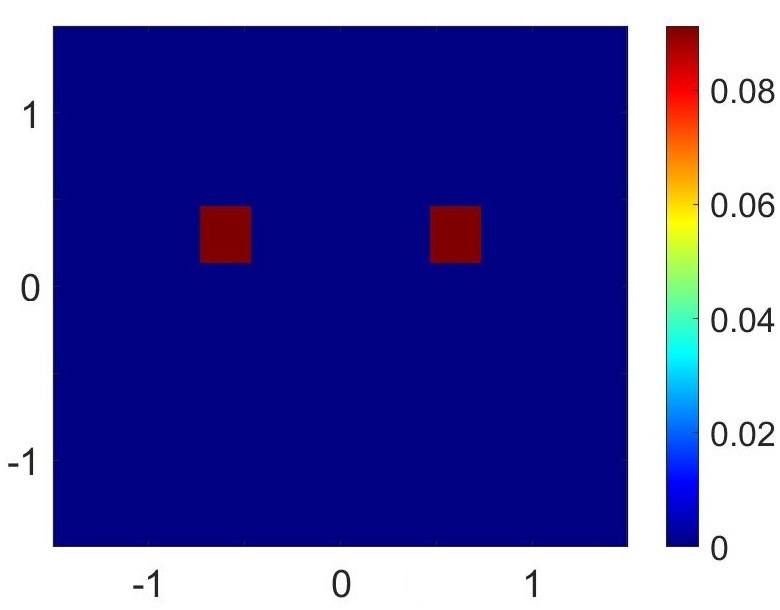}
		\subcaption{Ground Truth for setup shown in (a)}
	\end{subfigure}    
	\begin{subfigure}[t]{0.205\textwidth}
		\includegraphics[width=\textwidth]{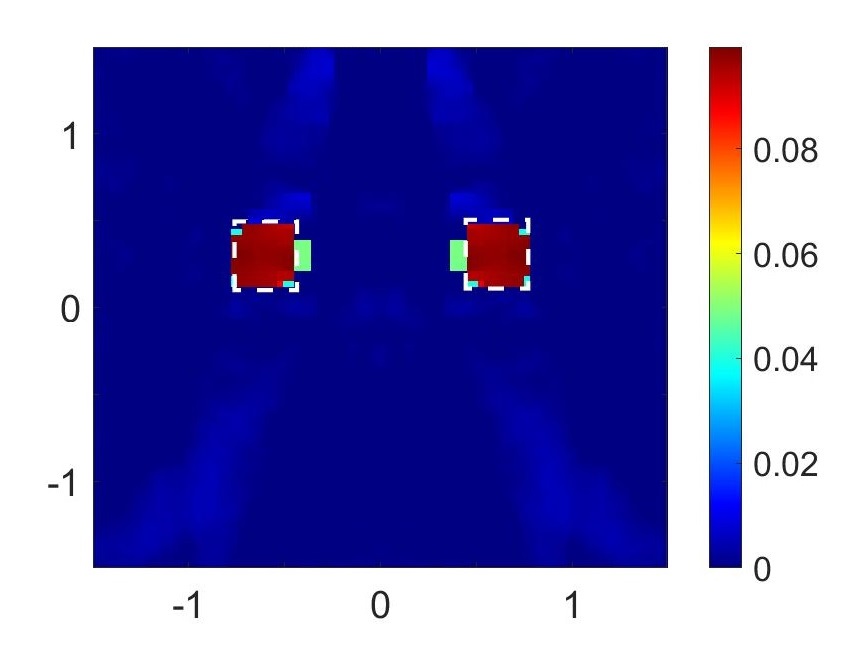}
		\subcaption{Reconstruction of $\operatorname*{Im}(\chi_{\text{RI}})$ using simulated measurements}
	\end{subfigure}
	\begin{subfigure}[t]{0.205\textwidth}
		\includegraphics[width=\textwidth]{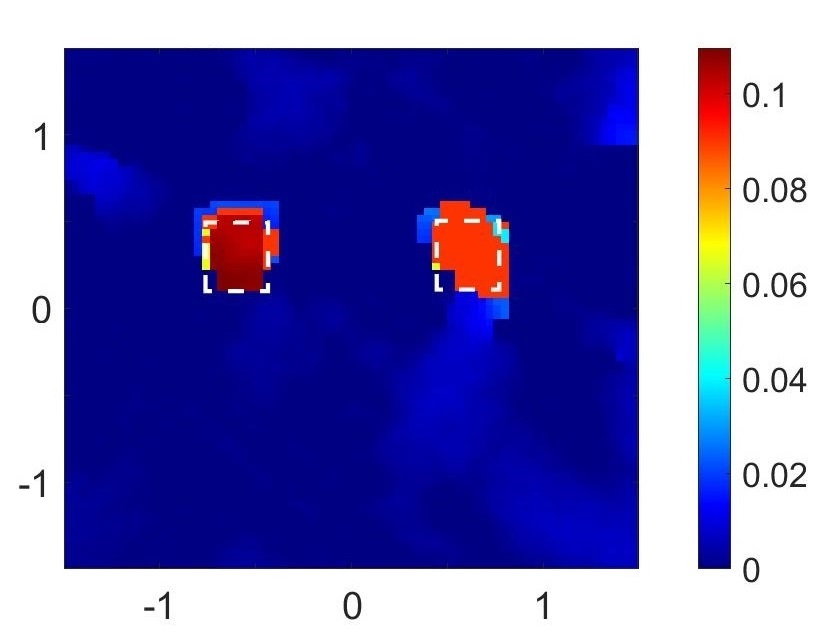}
		\subcaption{Reconstruction of $\operatorname*{Im}(\chi_{\text{RI}})$ using experimental measurements}
	\end{subfigure}	
	\caption{Experimental setup and results. (a) Experimental setup where two book stacks are placed on a Styrofoam platform. The cross section of each stack is $21 \times 30$ cm$^2$ and complex permittivity  $\epsilon_r = 3.4 + j 0.25$ ($\frac{2 \epsilon_I}{\pi} \sin^{-1}(1/\sqrt{\epsilon_R})=0.09$). (b) shows the ground truth profile of imaginary part of contrast $\operatorname*{Im}(\chi_{\text{RI}})$ representing the 2D cross section of experimental setup shown in (a). The (PSNR, SSIM) values for reconstructions in (c) and (d) are (28 dB, 0.874) and (22 dB, 0.88) respectively.}
	\label{LLHRI_exp1} 
	\vspace{-0.2\baselineskip}			
\end{figure*}

\begin{figure*}[h]
	\captionsetup[subfigure]{justification=centering}
	\centering
	\begin{subfigure}[t]{0.34\textwidth}
		\includegraphics[width=\textwidth]{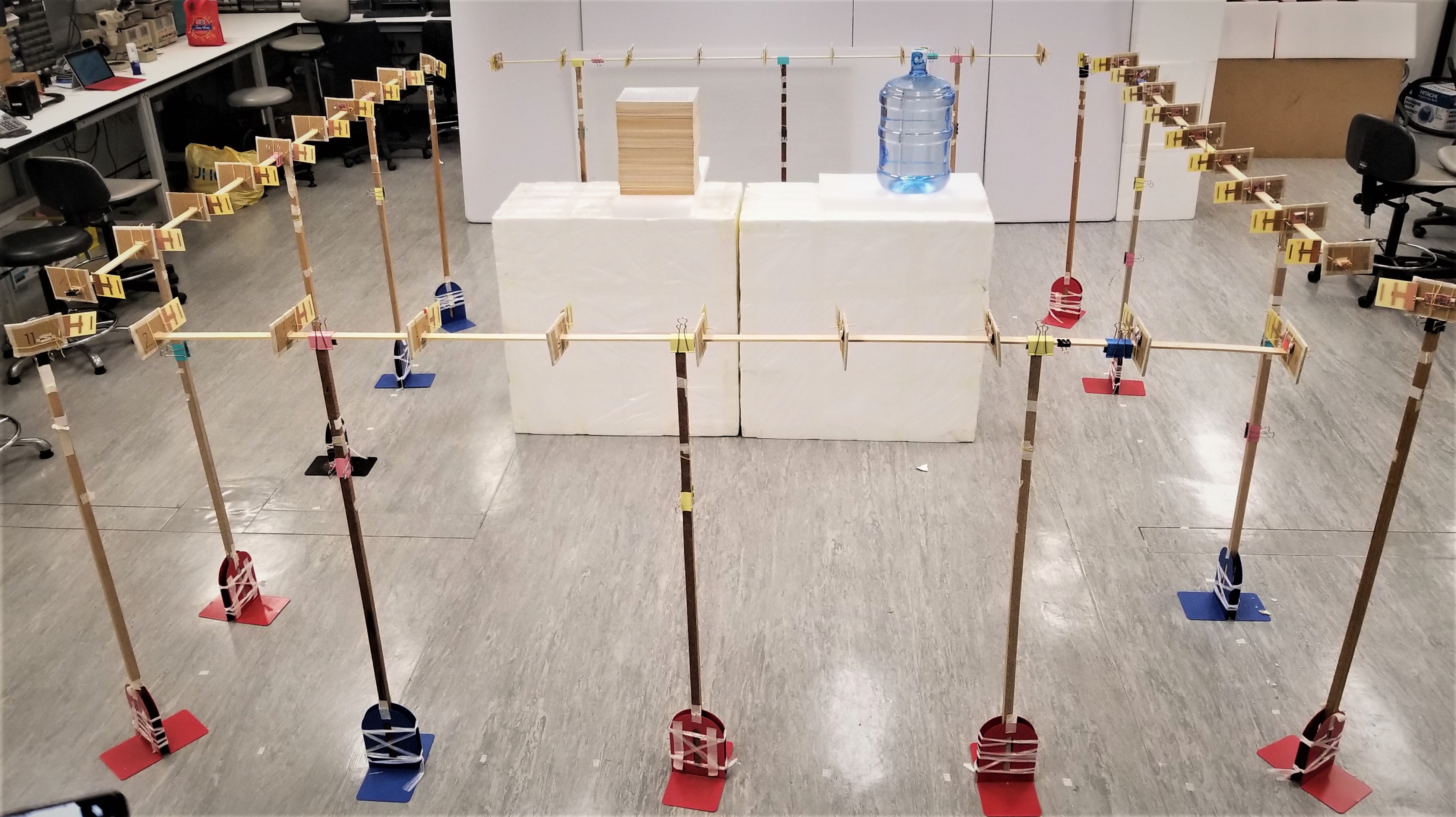}
		\subcaption{Experimental Setup}
	\end{subfigure}
	\begin{subfigure}[t]{0.2\textwidth}
		\includegraphics[width=\textwidth]{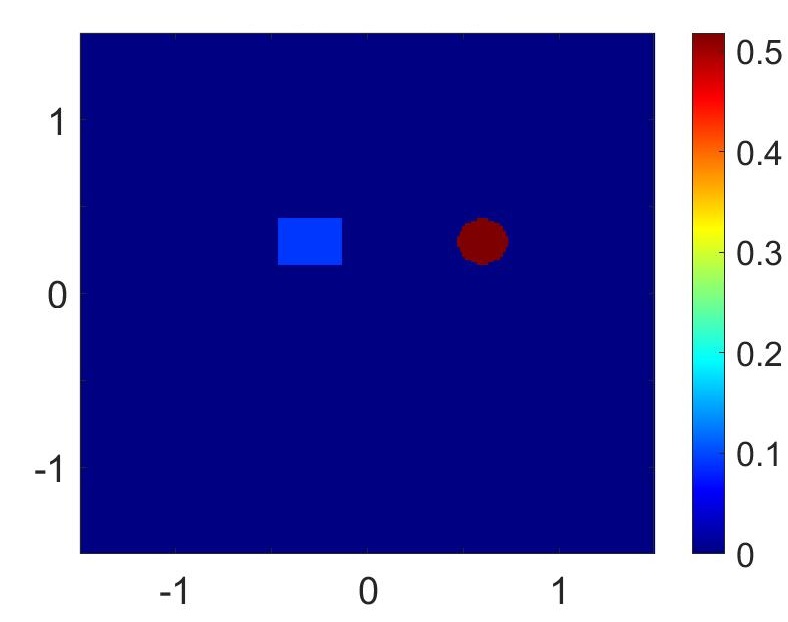}
		\subcaption{Ground Truth for setup shown in (a)}
	\end{subfigure}   
	\begin{subfigure}[t]{0.2\textwidth}
		\includegraphics[width=\textwidth]{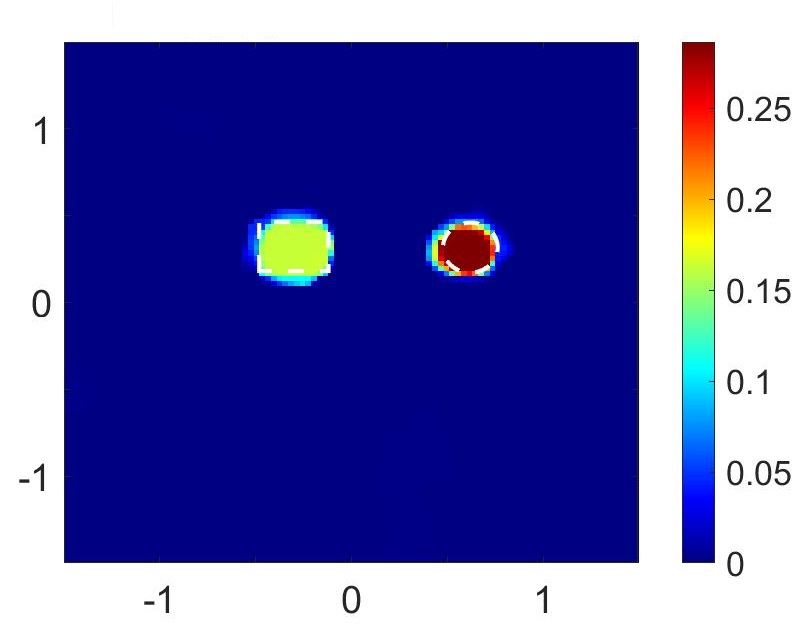}
		\subcaption{Reconstruction of $\operatorname*{Im}(\chi_{\text{RI}})$ using simulated measurements}
	\end{subfigure}	
	\begin{subfigure}[t]{0.205\textwidth}
		\includegraphics[width=\textwidth]{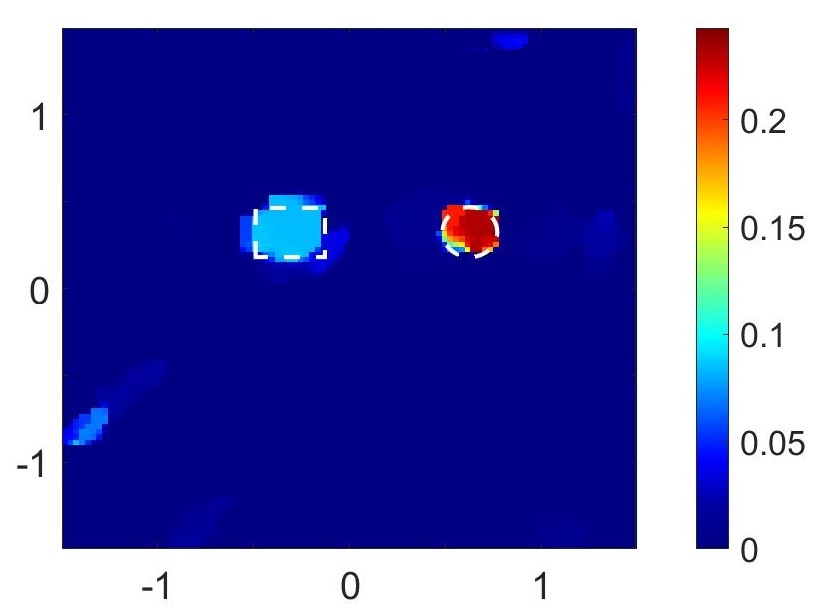}
		\subcaption{Reconstruction of $\operatorname*{Im}(\chi_{\text{RI}})$ using experimental measurements}
	\end{subfigure}	
	\caption{Experimental setup and results. (a) Experimental setup where a book stack and a circular container filled with water are placed on Styrofoam platform. The book stack act as medium permittivity object with $\epsilon_r= 3.4 + j 0.25$ ($\frac{2 \epsilon_I}{\pi} \sin^{-1}(1/\sqrt{\epsilon_R})=0.09$) and water-can act as very high permittivity object with $\epsilon_r= 77 + j 7.13$ ($\frac{2 \epsilon_I}{\pi} \sin^{-1}(1/\sqrt{\epsilon_R})=0.52$). (b) shows the ground truth profile of imaginary part of contrast $\operatorname*{Im}(\chi_{\text{RI}})$ representing the 2D cross section of experimental setup shown in (a). The (PSNR, SSIM) values for reconstructions in (c) and (d) are (25 dB, 0.874) and (27 dB, 0.88) respectively. }
	\label{LLHRI_exp2} 
	\vspace{-0.5\baselineskip}			
\end{figure*}
At height $d_h = 1.2$ m, the 2D cross sectional slice of 3D DOI in Fig. \ref{geometryexp} matches our simulation setup where DOI dimension and location of nodes are exactly the same as in Fig. \ref{geometry}. The difference between simulations and experiments arises in the measurement data, as the simulation setup is ideal 2D environment whereas the experimental setup is a 3D environment with 2D DOI at height 1.2 m from floor. Therefore, the experimental measurements contain distortions due to multipath reflections from the ceiling, floor and walls and other clutter outside the DOI. Furthermore, experimental data also contains errors and noise which is absent in the simulations. While the Yagi antennas reduce the effect of multiple reflections, distortions due to clutter will remain. The key technique which can handle these distortions and error in the experimental setup is the background subtraction framework (\ref{Eq_RIdBTBS}).

Fig. \ref{LLHRI_exp1}(a) details our first experiment where two book stacks are placed upon the Styrofoam platform. Our goal is to image these book stacks and the cross section of each stack is $21 \times 30$ cm$^2$. Fig. \ref{LLHRI_exp1}(b) shows the ground truth profile of the imaginary part of the contrast function $\operatorname*{Im}(\chi_{\text{RI}})$ representing the 2D cross section of the experimental setup shown in Fig. \ref{LLHRI_exp1}(a). The complex relative permittivity of the papers used in the books is $\epsilon_r = 3.4 + j 0.25$ (estimated using a cavity resonator). 

Fig. \ref{LLHRI_exp1}(c) shows the numerical reconstruction result where the measurements are generated numerically using the profile shown in Fig. \ref{LLHRI_exp1}(b). It can be seen that the reconstruction result using synthetic data is accurate. To achieve the same performance with experimental data, we have to use background subtraction. Therefore in the experiments, we first take measurements without book stacks (to capture all multipath reflections and clutter from the background) and then take another measurement with book stacks. To this end, at initial time instant, $t_0$, we take RSSI measurements  $P^{t_0}$ when there is only Styrofoam inside the DOI but no book stacks. We then place the book stacks as shown at time instant $t_0+\Delta t$ and take the RSSI measurements $P^{t_0+\Delta t}$ again. This set of RSSI data will contain scattering from the books along with all the scattering from the background (which was included in $P^{t_0}$). We then use xPRA-LM (\ref{Eq_RIdBTBS}) to estimate change in contrast profile due to the placement of the book stacks.

It can be seen that the experimental result (Fig. \ref{LLHRI_exp1}(d)) obtained is on par with the simulation result (Fig. \ref{LLHRI_exp1}(c)) and show very good accuracy in reconstruction. This is despite the fact that the experimental setup is not an ideal 2D setup and it contains significant multipath reflections from clutter inside and outside the DOI. This can be credited to the ability of xPRA-LM to perform background subtraction  (\ref{Eq_RIdBTBS}) (and is also demonstrated in simulation in Fig. \ref{LLHRI_TBS}). 

Another experimental setup is shown in Fig. \ref{LLHRI_exp2}(a) in which, there is a book stack and a container of water placed inside DOI. The book stack has a rectangular cross section of $30\times 21$ cm$^2$ and the container of water has a circular cross section with diameter $26$ cm. The book stack acts as a strong scatterer with permittivity $\epsilon_r= 3.4 + j 0.25$ and the container of water acts as a very strong scatterer with permittivity $\epsilon_r= 77 + j 7$ at 2.4 GHz. The 2D cross section of the ground truth of $\operatorname*{Im}(\chi_{\text{RI}})$ is shown in Fig. \ref{LLHRI_exp2}(b). Similarly to the previous results, the reconstruction using experimental data (Fig. \ref{LLHRI_exp2}(d)) matches the reconstruction using simulation data (Fig. \ref{LLHRI_exp2}(c)) and provides accurate imaging of the two objects with different permittivity. Both the PSNR and SSIM are again good. In contrast the results shown in Appendix B have PSNR and SSIM values that are both very poor being less than 10 dB and 0.8 respectively.

\section{Conclusion}
In this paper we have presented a phaseless inverse scattering technique, xPRA-LM. It is demonstrated by both simulations and experiments that  xPRA-LM can provide accurate reconstructions of the imaginary component of the contrast function under a wide variety of scattering conditions including strong scattering (when scatterer has large size and high permittivity). The xPRA-LM method is based on corrections to the Rytov approximation which we obtain by approximating the gradient of the scattered field inside the scatterer using high frequency approximations in lossy media. We also incorporate background subtraction which helps to enhance experimental performance and remove distortions caused by background scattering or clutter.
 
We demonstrate the performance of the proposed xPRA-LM method for the use-case of indoor imaging using phaseless WiFi measurements and provide extensive simulation and experimental results, covering a wide range of scenarios including weak scattering and extremely strong scattering. The experiments are performed in an indoor environment with clutter inside as well as outside the DOI. We demonstrate that the approach provides accurate reconstruction (of both shape and amplitude) up to relative permittivities of $15+j1.5$ while maintaining accurate shape reconstruction up to relative permittivities  $\epsilon_r=77+j 7$ where the electrical size of the scatterer is greater than $20 \lambda$. To the best of our knowledge, this is the first inverse scattering approach for indoor imaging which provides accurate shape reconstruction of the scatterers in DOI along-with information about their complex permittivity even under extremely strong scattering conditions. Furthermore the technique requires only phaseless measurements. More generally our technique is applicable to a wide variety of RF imaging applications and has the potential to open up this area.

\appendices
\section{}
\label{MultScattering}
Ray equations in (\ref{Eq_complexwavevectorfield}) (and (\ref{Eq_raytraced1})) represent the first order event of reflection-refraction-absorption. There will also be higher order events inside a finite size scatterer even if it is larger than or comparable to $\lambda_0$. For lossy scatterers, such analysis of higher order interaction is extremely difficult. There have been limited attempts made to partially model such higher order events under certain practical constraints \cite{yang2009effective, groth2016numerical, zhang2015refractive}. For example, recent attempts \cite{yang2009effective, groth2016numerical} have been made to trace rays inside a large polygon shaped lossy scatterer (lossy hexagonal ice crystals) but in general predicting the behavior of the higher order rays inside a lossy scatterer is still an open problem \cite{groth2016numerical}. Fortunately, previous numerical analysis \cite{yang2009effective} shows that for a lossy scatterer the higher order scattering events inside the scatterer do not have a significant effect on the value of the field scattered by the scatterer. This can be further justified using physical intuition where due to loss or absorption, the energy carried by a higher order ray will be less than a lower order ray and the energy will keep decreasing in the higher order rays depending on the size and absorbing properties of the object. In such case, the scattered field due to the object will be dominated by the first order event. 

In our configuration even though we are assuming objects to be low-loss, it only implies $\epsilon_R\gg \epsilon_I$ and doesn't imply absorption ($\epsilon_I$) is negligible. For example, in the case of water (or human body), $\epsilon_R = 77, \epsilon_I=7$. The absorption associated with large values of $\epsilon_I=7.13$ is substantial. It is therefore justifiable in our configuration to use only the first order ray in (\ref{Eq_raytraced1}).

Assume at any given point, the net field inside a scatterer (which exhibits substantial absorption) is the superposition of $N$ higher order events. We can rewrite (\ref{Eq_rytHFtotal}) by adding the higher order rays to the first order ray (\ref{Eq_raytraced1}),
\begin{equation}
	\label{Eq_rytHFtotalHO}
	\begin{aligned}
		E_i  (\bm{r})  e^{jk_0\phi_s^H(\bm{r})} =  \underbrace{A_t e^{- \alpha_t} e^{j \beta_t} }_\text{\scriptsize{first order ray given in (\ref{Eq_raytraced1})}} + \underbrace{ \sum_{n=1}^{N} A_{n} e^{- \alpha_n} e^{j  \beta_n} }_\text{\scriptsize{Higher order rays}}
	\end{aligned}
\end{equation}
where $\phi_s^H$ is the phase of the scattered field after considering higher order rays (whereas $\phi_s$ in (\ref{Eq_rytHFtotal1}) is derived by only considering only the first order ray). The first order transmitted ray and its phase $\beta_t$ and attenuation $\alpha_t$ parameters are already defined in (\ref{Eq_rytHFtotal}), and are a function of the effective refractive index $V_R$ and $V_I$. The phase $\beta_n$ and attenuation $\alpha_n$ parameters in the higher order rays depend on the effective refractive index seen by the higher order rays. It has been shown numerically \cite{yang2009effective} that the effective refractive index for the higher order rays is very close to that seen by the lower order rays. This  implies that $|\alpha_n| \approx |\alpha_t|$ and $|\beta_n| \approx |\beta_t|$. Furthermore, for a lossy scatterer of large size, the energy in the higher order rays will keep decreasing and if the scatterer exhibits substantial loss (due to large $\epsilon_I$ and large size), we can say that the higher order rays will have much less energy than the first order ray and hence, $A_n \ll A_t$. Noting these observations and following the next step  (\ref{Eq_rytHFtotal1}) to find $\phi_s$, we get
\begin{equation}
	\label{Eq_rytHFtotal1HO}
	\begin{aligned}
	 &\phi_s^H(\bm{r}) = &\frac{1}{jk_0 }\text{ln}\biggl[(\frac{A_t}{A_0} e^{-\alpha_t} e^{j \beta_t} + \sum_{n=1}^{N} \frac{A_n}{A_0} e^{-\alpha_n} e^{j \beta_n} )e^{-j k_0 \bm{\hat{k}_i \cdot r}} \biggr]
	\end{aligned}
\end{equation}
which can be rewritten as,
\begin{equation}
	\label{Eq_rytHFtotal1HO1}
	\begin{aligned}
		&\phi_s^H(\bm{r}) \\= &\frac{1}{jk_0 }\text{ln}\biggl[\frac{A_t}{A_0} e^{-\alpha_t} e^{j \beta_t} e^{-j k_0 \bm{\hat{k}_i \cdot r}} (1 + \sum_{n=1}^{N} \frac{A_n}{ A_t} e^{-\Delta \alpha_n} e^{j \Delta \beta_n} ) \biggr]
		\\=
		&\frac{1}{jk_0 }\text{ln}\biggl[\frac{A_t}{A_0} e^{-\alpha_t} e^{j \beta_t} e^{-j k_0 \bm{\hat{k}_i \cdot r}}\biggr] 
		\\  & \qquad \qquad \qquad +\frac{1}{jk_0 }\text{ln} \biggl[1 + \sum_{n=1}^{N} \underbrace{\frac{A_n}{ A_t} e^{-\Delta \alpha_n} e^{j \Delta \beta_n} }_{\text{denote as } x}\biggr].
	\end{aligned}
\end{equation}
We know that $A_n\ll A_t$ and hence we can say $x \ll 1$ and hence $\ln(1+x) \approx 0$. Using this, we can neglect the second term in (\ref{Eq_rytHFtotal1HO1}). The remaining equation shows that $\phi^H_s$ only predominantly depends on the first order ray and the incident ray $\phi^H_s \approx \phi_s$ where $\phi_s$ is given in (\ref{Eq_rytHFtotal1}). It is important to note that even if absorption inside the scatterer is small, $\phi_s^H$ will still predominantly depend on the first order ray because even for small loss, $A_n <  A_t$. However, this will cause second order errors in the result, which may lead to small distortions in the reconstructions. 

\section{Real Part of Reconstructions}
\label{realpart}
This appendix provides the real part of the reconstructions of the contrast function for a selection of the configurations used in the simulation section of the paper. The reconstruction results are obtained from $\chi_{\text{RI}}(\bm{r})$ as $\frac{(2+\operatorname{Re}(\chi_{\text{RI}}(\bm{r})))^2}{4}$ (using (\ref{Eq_rytovfulldB6-1})). The ground truth is taken as the actual $\epsilon_R$. These results are provided to show that the reconstruction of the real part of the contrast function only provides good results for weak scattering. In addition, these results are inline with the multitude of previous results for the real part \cite{murch1990inverse, bates1976extended, wu2003wave} providing a benchmark for comparison. It also allows us to compare the results with our reconstructions of the imaginary component. This demonstrates the enormous benefit of using the imaginary component of the contrast function. In particular, in the results shown in this appendix, when the permittivity deviates from unity, there is significant distortion in the real part of the contrast function. This clearly shows that reconstructions of the real part of the contrast function using RA are of virtually of no use in the applications we are considering. Apart from the weak scattering example all results have PSNR and SSIM values that are poor being less than 10 dB and 0.8 respectively.
\begin{figure}[!h]
	\captionsetup[subfigure]{justification=centering}
		\centering
		\begin{subfigure}[t]{0.21\textwidth}
			\includegraphics[width=\textwidth]{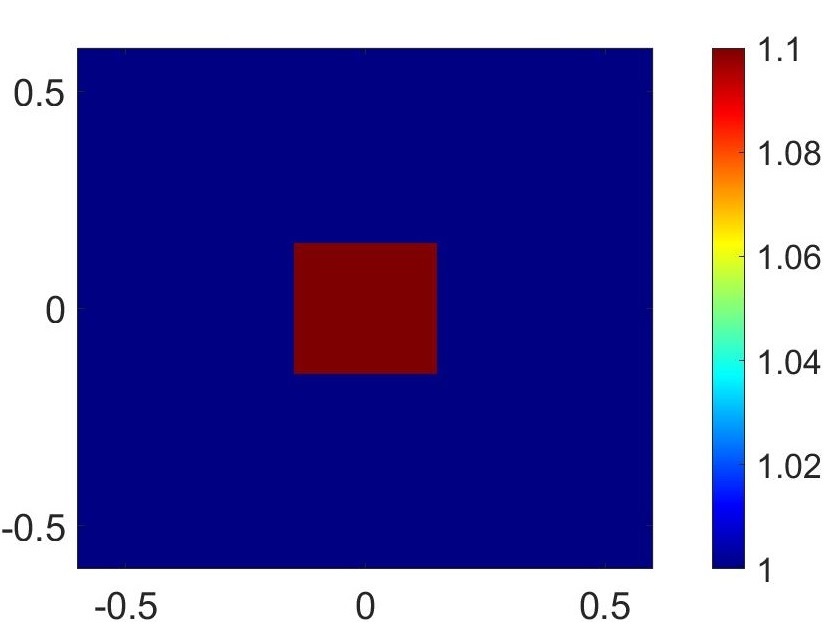}
			\subcaption{Ground Truth: $\epsilon_R = 1.10, \epsilon_I=0.11$}
		\end{subfigure}       
		\begin{subfigure}[t]{0.21\textwidth}
			\includegraphics[width=\textwidth]{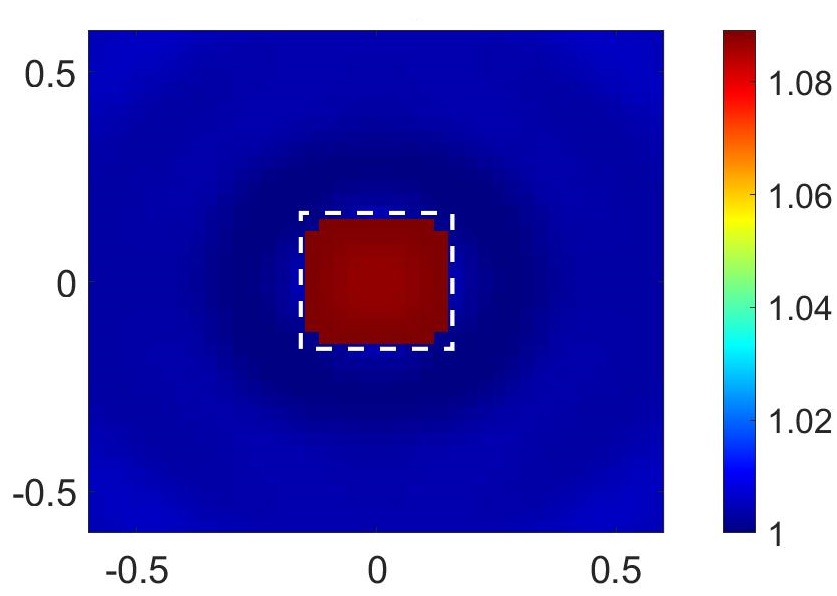}
			\subcaption{{Reconstruction of ${\epsilon_R}$}}
		\end{subfigure}       	
		\caption{Real part reconstructions for profile shown in Fig. \ref{LLLRI} (PSNR = 48 dB, SSIM = 0.96).}
		\label{LLLRI_R} 
		\vspace{-0.5\baselineskip}	
\end{figure}
\begin{figure}[!h]
	\captionsetup[subfigure]{justification=centering}
		\centering
		\begin{subfigure}[t]{0.20\textwidth}
			\includegraphics[width=\textwidth]{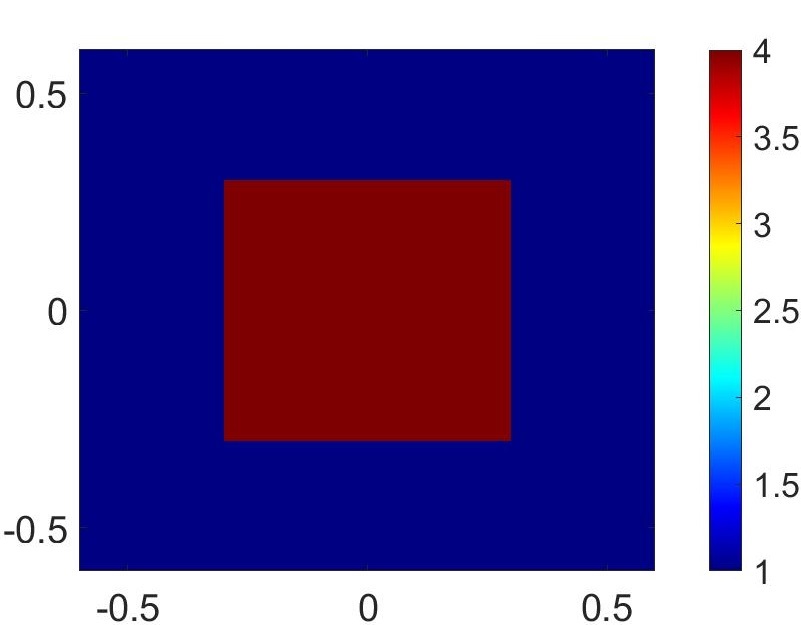}
			\subcaption{Ground Truth: $\epsilon_R = 4, \epsilon_I=0.4$}
		\end{subfigure}       
		\begin{subfigure}[t]{0.208\textwidth}
			\includegraphics[width=\textwidth]{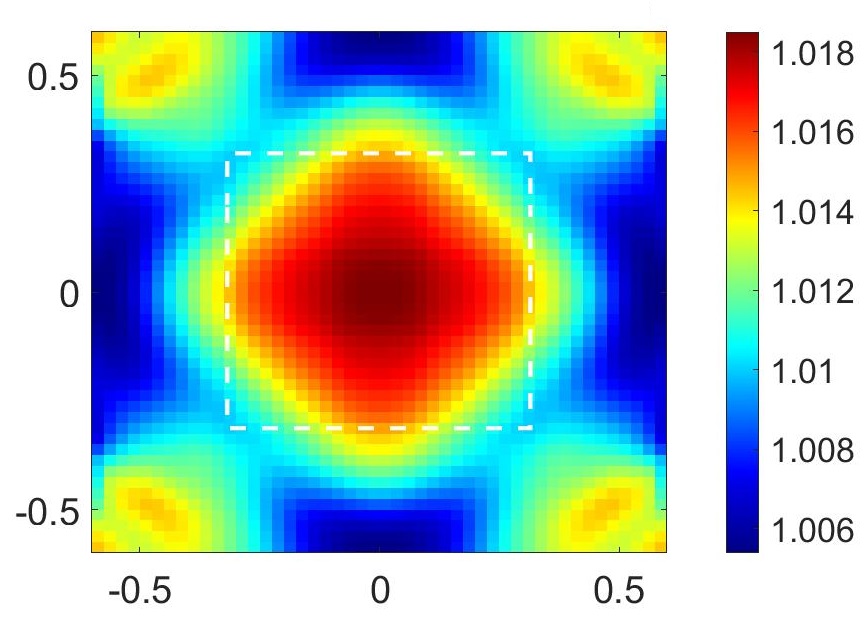}
			\subcaption{{Reconstruction of ${\epsilon_R}$}}
		\end{subfigure}	
		\caption{Real part reconstructions for profile shown in Fig. \ref{LLHRIbooks1} (PSNR = 1.7 dB, SSIM=0.77).}
		\label{LLHRIbooks1_R} 
		\vspace{-0.5\baselineskip}	
\end{figure}
\begin{figure}[!h]
	\captionsetup[subfigure]{justification=centering}
		\centering
		\begin{subfigure}[t]{0.20\textwidth}
			\includegraphics[width=\textwidth]{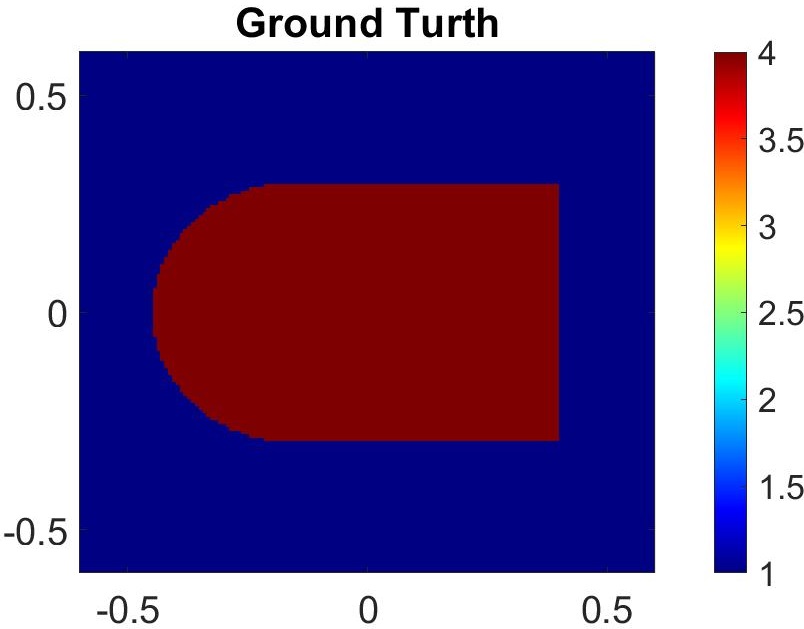}
			\subcaption{Ground Truth: $ \epsilon_R=4, \epsilon_I=0.4$}
		\end{subfigure}       
		\begin{subfigure}[t]{0.20\textwidth}
			\includegraphics[width=\textwidth]{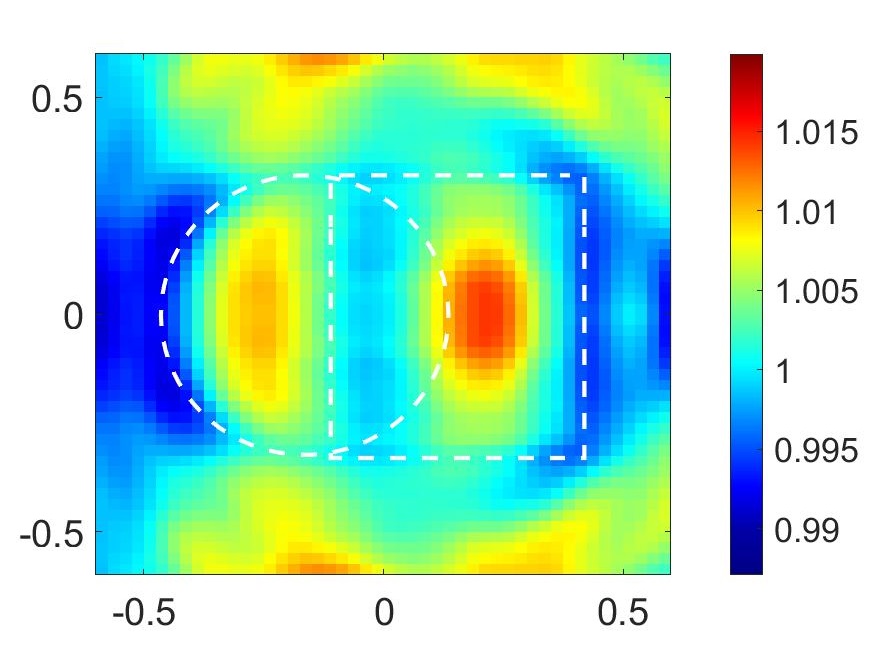}
			\subcaption{{Reconstruction of ${\epsilon_R}$}}
		\end{subfigure}       
		\caption{Real part reconstructions for profile shown in Fig. \ref{LLHRIbooks2} (PSNR = 0.7 dB, SSIM=0.58).}
		\label{LLHRIbooks2_R} 
		\vspace{-0.6\baselineskip}	
\end{figure}
\begin{figure}[!h]
	\captionsetup[subfigure]{justification=centering}
		\centering
		\begin{subfigure}[t]{0.20\textwidth}
			\includegraphics[width=\textwidth]{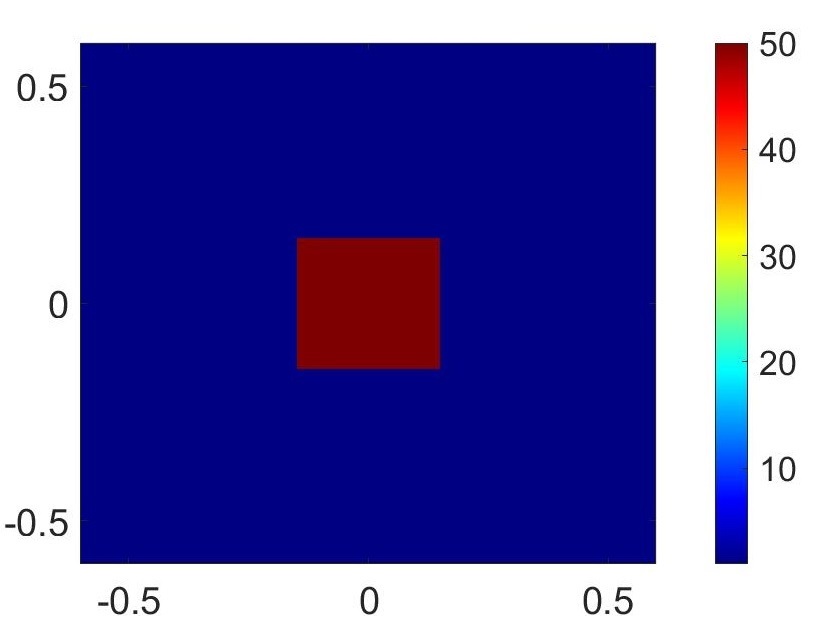}
			\subcaption{Ground Truth with ${\epsilon_R} = 50, \epsilon_I=5$}
		\end{subfigure}       
		\begin{subfigure}[t]{0.207\textwidth}
			\includegraphics[width=\textwidth]{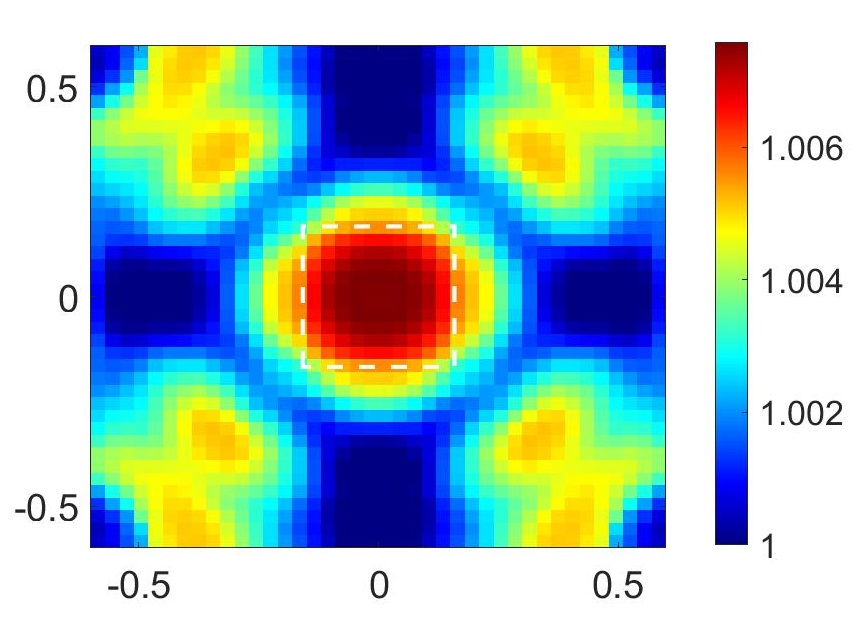}
			\subcaption{{Reconstruction of ${\epsilon_R}$}}
		\end{subfigure}       
		\caption{Real part reconstructions for profile shown in Fig. \ref{LLHRIhuman} (PSNR = 2 dB, SSIM=0.73).}
		\label{LLHRIhuman_R} 
		\vspace{-0.5\baselineskip}	
\end{figure}	
\begin{figure}[!h]
		\captionsetup[subfigure]{justification=centering}
		\centering
		\begin{subfigure}[t]{0.20\textwidth}
			\includegraphics[width=\textwidth]{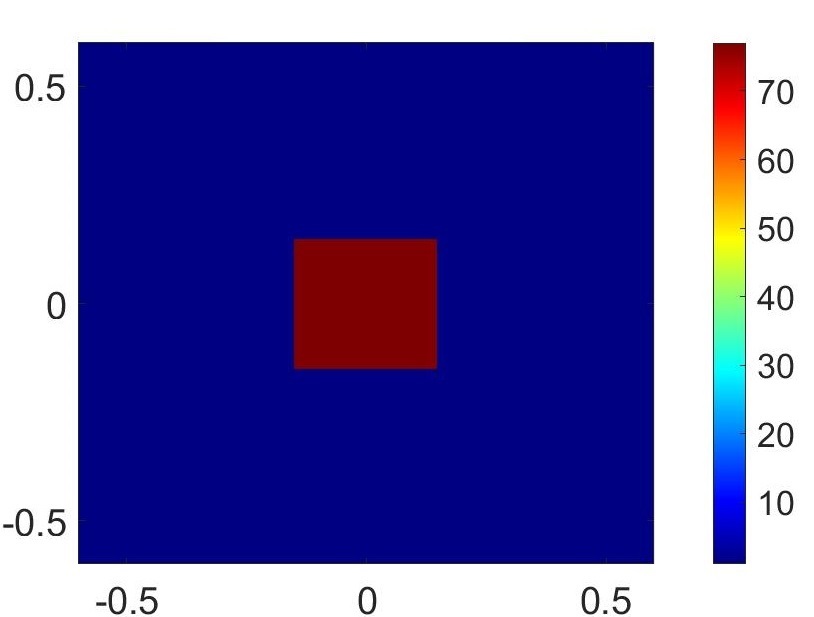}
			\subcaption{Ground Truth with $\epsilon_R = 77, \epsilon_I=7.7$}
		\end{subfigure}       
		\begin{subfigure}[t]{0.207\textwidth}
			\includegraphics[width=\textwidth]{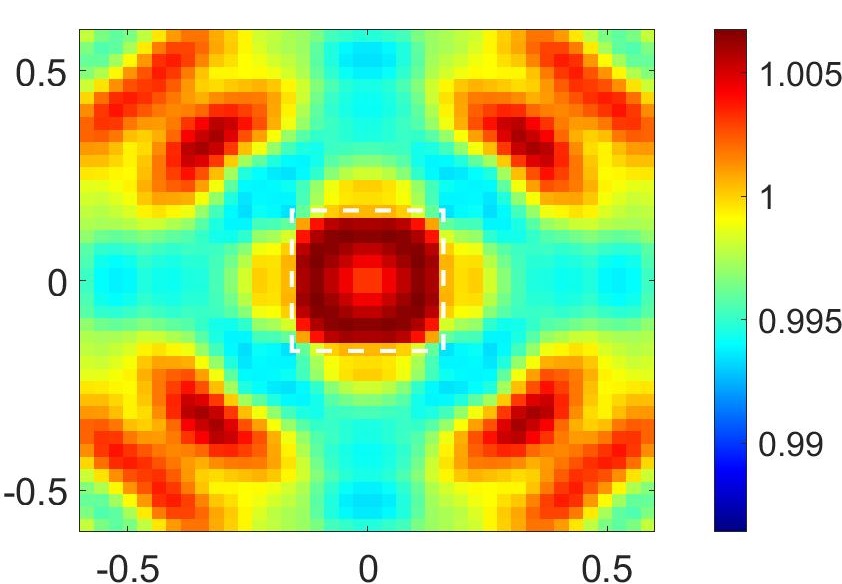}
			\subcaption{{Reconstruction of ${\epsilon_R}$}}
		\end{subfigure}       
		\caption{Real part reconstructions for profile shown in Fig. \ref{LLHRIwater}  (PSNR = 0.8 dB, SSIM=0.61).}
		\label{LLHRIwater_R} 
		\vspace{-0.2\baselineskip}
\end{figure}
\begin{figure}[!h]
		\captionsetup[subfigure]{justification=centering}
		\centering
		\begin{subfigure}[t]{0.20\textwidth}
			\includegraphics[width=\textwidth]{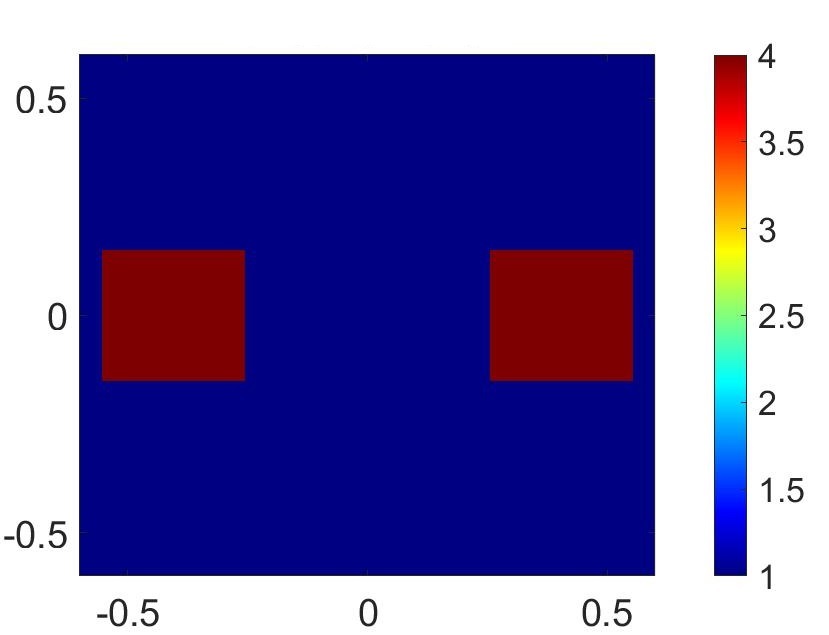}
			\subcaption{Ground Truth with $\epsilon_R = 4, \epsilon_I=0.4$}
		\end{subfigure}       
		\begin{subfigure}[t]{0.207\textwidth}
			\includegraphics[width=\textwidth]{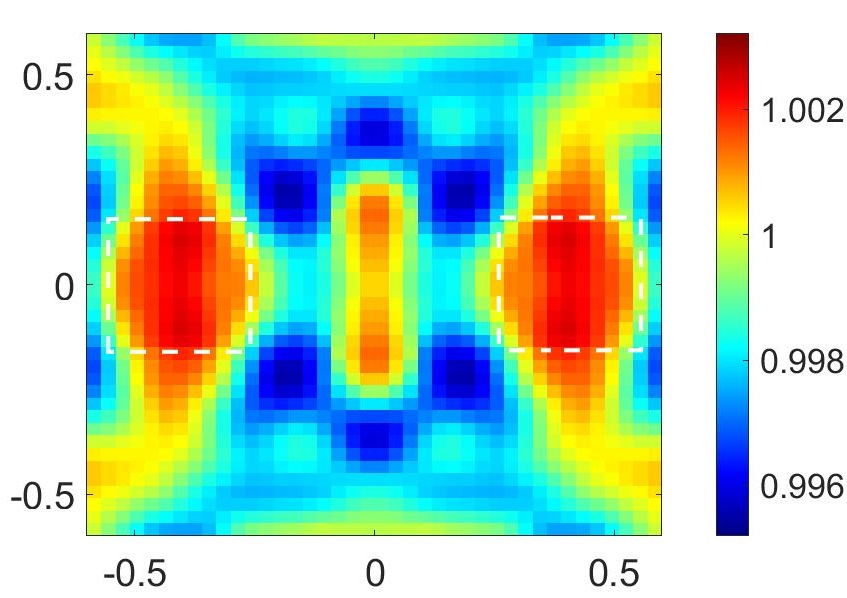}
			\subcaption{{Reconstruction of ${\epsilon_R}$}}
		\end{subfigure}       
		\caption{Real part reconstructions for profile shown in Fig. \ref{LLHRI_multiple1} (PSNR = 0.5 dB, SSIM=0.63).}
		\label{LLHRI_multiple1_R} 
		\vspace{-0.2\baselineskip}			
\end{figure}
\begin{figure}[!h]
		\captionsetup[subfigure]{justification=centering}
		\centering
		\begin{subfigure}[t]{0.20\textwidth}
			\includegraphics[width=\textwidth]{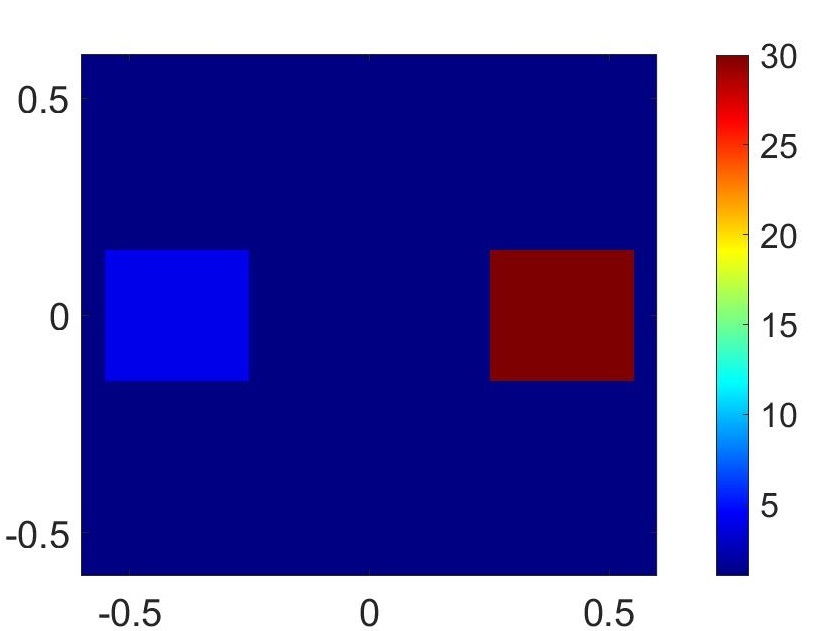}
			\subcaption{Ground Truth: $\epsilon_R = 30, \epsilon_I=3$ (right object) \& $\epsilon_R = 4, \epsilon_I=0.4$ (left object)}
		\end{subfigure}       
		\begin{subfigure}[t]{0.207\textwidth}
			\includegraphics[width=\textwidth]{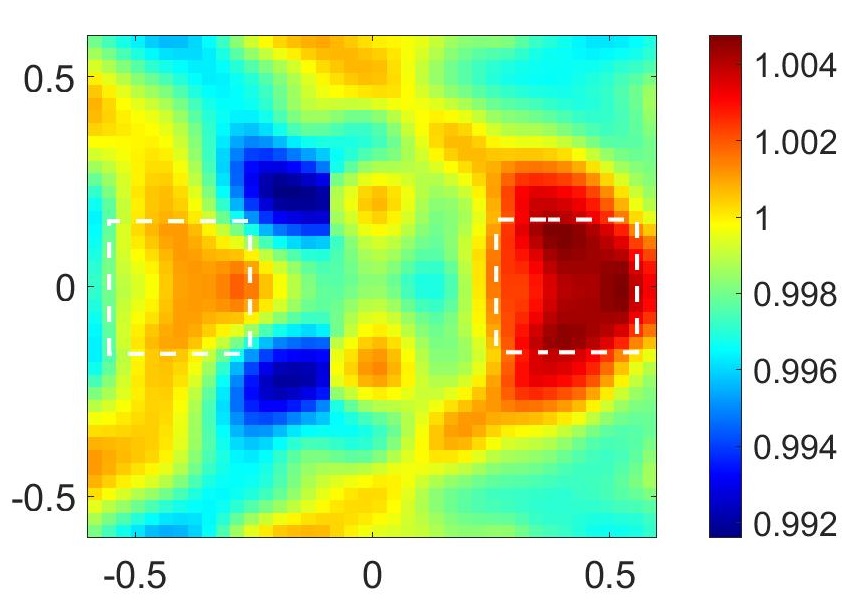}
			\subcaption{{Reconstruction of ${\epsilon_R}$}}
		\end{subfigure}       
		\caption{Real part reconstructions for profile shown in Fig. \ref{LLHRI_multiple2}(PSNR = 0.2 dB, SSIM=0.65).}
		\label{LLHRI_multiple2_R} 
		\vspace{-0.5\baselineskip}	
\end{figure}

\bibliographystyle{IEEEtran}
\bibliography{VTref}
%
\end{document}